\documentclass[12pt]{article}
\usepackage{jheppub}
\pdfoutput=1
\usepackage{graphicx}
\usepackage{amsmath}
\usepackage{amssymb}
\usepackage{amsthm}
\usepackage{mathrsfs}
\usepackage{subfigure}
\usepackage{bm}
\usepackage{dsfont}
\usepackage{comment}
\usepackage[utf8]{inputenc}
\usepackage{calc}
\usepackage{accents}
\usepackage{amsmath}
\usepackage{amssymb}
\usepackage{amsfonts}
\newtheorem{definition}{Definition}
\usepackage{braket}
\usepackage[normalem]{ulem}
\usepackage{color}
\usepackage{bbold}
\usepackage{ulem}
\usepackage{mathtools}
\usepackage{tensor} 
\usepackage{tikz}

\title{The Structure of Quantum Singularities on a Cauchy Horizon}

\author{Arvin Shahbazi-Moghaddam}

\affiliation{Center for Theoretical Physics and Department of Physics,\\
University of California, Berkeley, CA 94720, U.S.A.}

\emailAdd{arvinshm@gmail.com}

\abstract{Spacetime singularities pose a long-standing puzzle in quantum gravity. Unlike Schwarzschild, a generic family of black holes gives rise to a Cauchy horizon on which, even in the Hartle-Hawking state, quantum observables such as $\langle T_{\mu\nu} \rangle$ -- the expectation value of the stress-energy tensor -- can diverge, causing a breakdown of semiclassical gravity. Because they are diagnosed within quantum field theory (QFT) on a smooth background, these singularities may provide a better-controlled version of the spacetime singularity problem, and merit further study.

Here, I highlight a mildness puzzle of Cauchy horizon singularities: the $\langle T_{\mu\nu} \rangle$ singularity is significantly milder than expected from symmetry and dimensional analysis. I address the puzzle in a simple spacetime $\mathcal{W}_P$, which arises universally near all black hole Cauchy horizons: the past of a codimension-two spacelike plane in flat spacetime. Specifically, I propose an extremely broad QFT construction in which, roughly speaking, Cauchy horizon singularities originate from operator insertions in the causal complement of the spacetime. The construction reproduces well-known outer horizon singularities (e.g., in the Boulware state), and remarkably, when applied to $\mathcal{W}_P$, gives rise to a universal mild singularity structure for \emph{robust singularities}, ones whose leading singular behavior is state-independent. I make non-trivial predictions for \emph{all} black hole Cauchy horizon singularities using this, and discuss extending the results beyond robust singularities and the strict near Cauchy horizon limit.}

\begin{document}

\maketitle
\flushbottom

\section{Introduction}

Black holes are a major gateway into exploring quantum gravity. Spacetime singularities are a particular way in which black holes bring out the quantum nature of spacetime. A characteristic of singularities is the divergence of fields in a way which breaks the semiclassical approximation. The study of singularities in quantum gravity ought to involve, at the very least, the singularity structure, i.e., which observables diverge and with what degree. While the $r=0$ moment in the Schwarzschild metric is often discussed as a standard example of a singularity, this is not a generic case in stationary black holes, at least for a broad class of theories. Instead, it is known that generic stationary black holes, and small perturbations of them, have Cauchy horizons.

\begin{figure}[htbp]
\centering
\begin{minipage}{0.34\textwidth}
    \includegraphics[width=\linewidth]{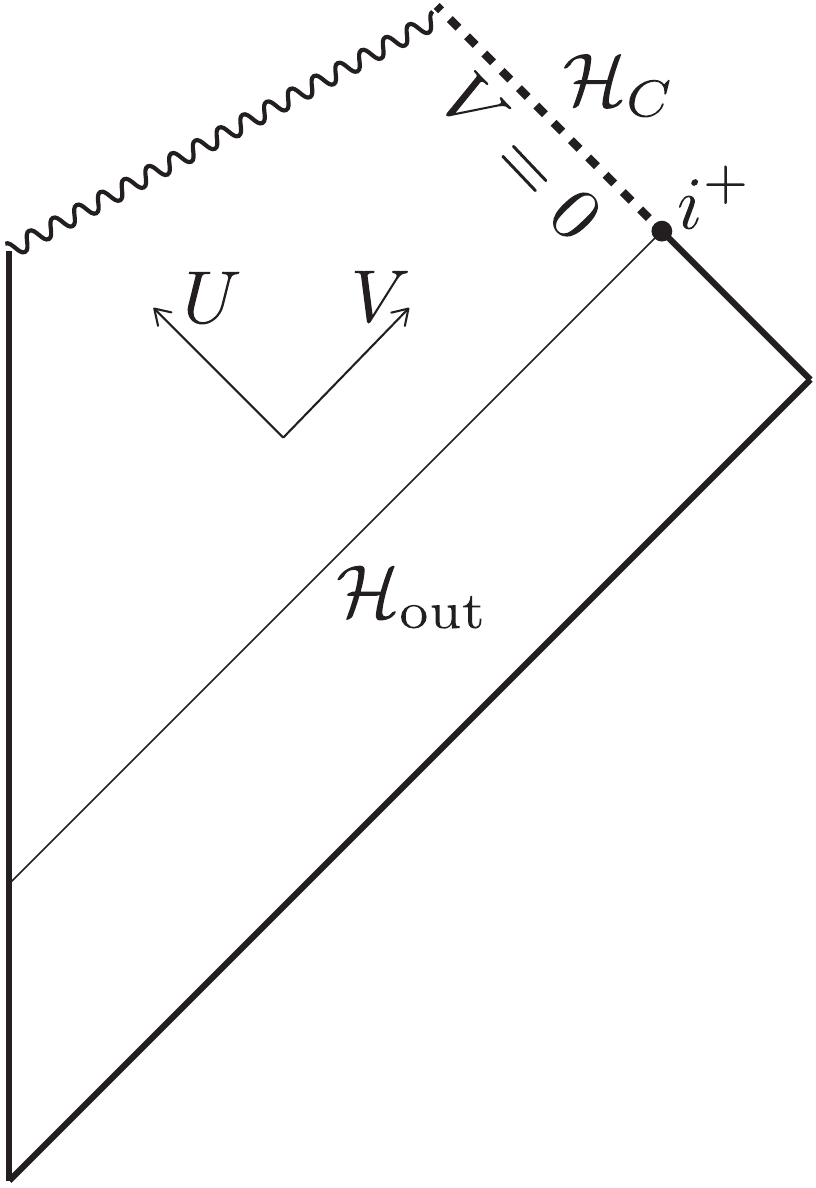} 
\end{minipage}
\hfill
\begin{minipage}{0.48\textwidth}
    \includegraphics[width=\linewidth]{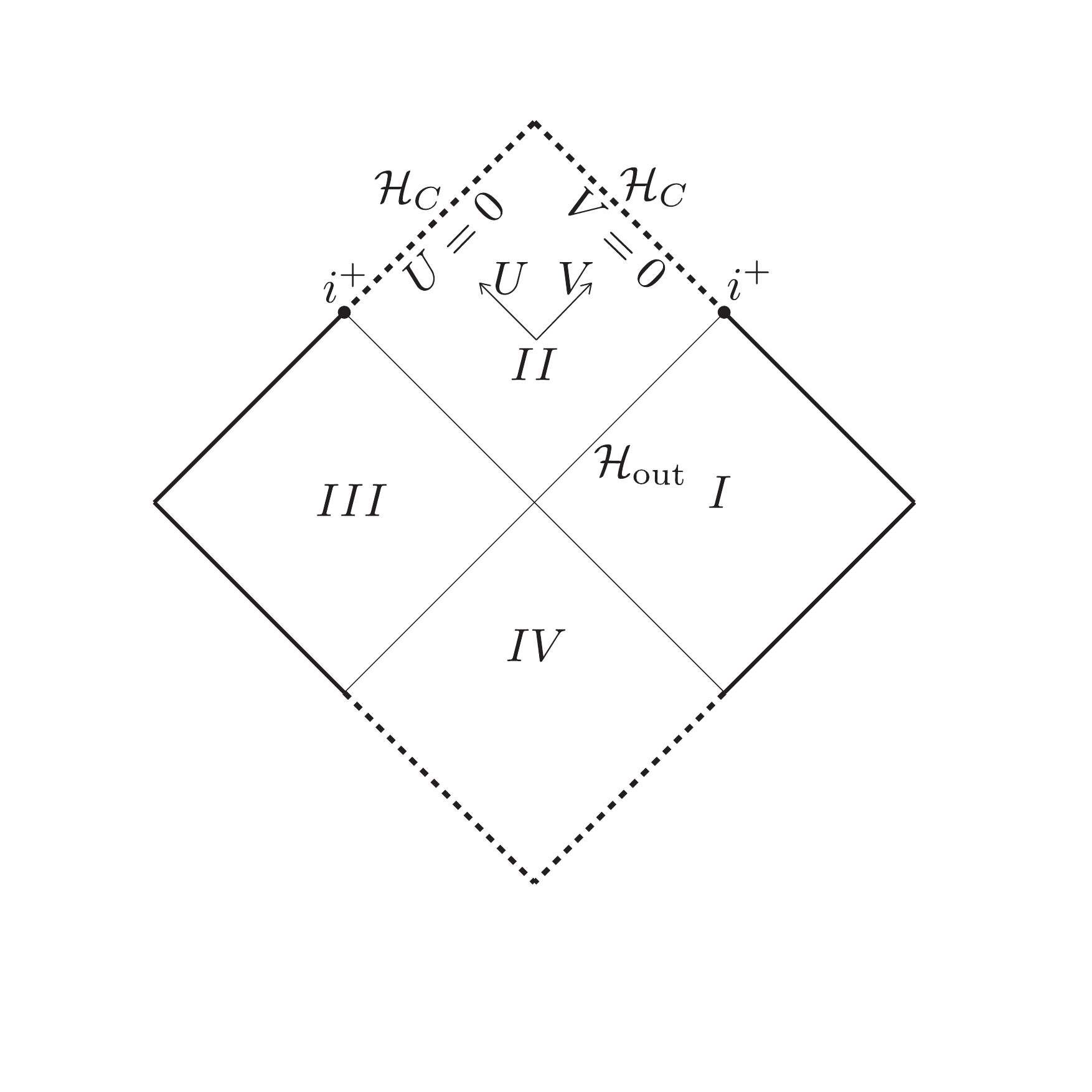} 
\end{minipage}
\caption{Black hole spacetimes with Cauchy horizons $\mathcal{H}_C$. On the left, a charged and/or rotating black hole formed by collapse in asymptotically flat spacetime. If the outer horizon settles down to stationarity appropriately, a Cauchy horizon (dashed line at $V=0$) would emanate from timelike infinity ($i^+$), on which the quantum state can have observable singularities, e.g., in the expectation value of the stress-energy tensor. On the right, a maximally extended black hole spacetime with a Cauchy horizon ($V=0$ and $U=0$ dotted lines) is shown. The Cauchy horizon is stable under small classical perturbations and can have quantum singularities. Regions I and III are the exteriors, region II and IV are the black hole and white hole regions respectively. On the unperturbed background, the HH state can be defined on the entire diagram, while the Unruh state can be defined in region I $\cup$ II.}
\label{fig1}
\end{figure}

Let $\mathcal{M}$ be a globally hyperbolic spacetime resulting from the maximal Cauchy evolution of a complete initial data slice.\footnote{By a complete initial data slice, we mean one which cannot be embedded inside a larger one.} If $\mathcal{M}$ is extendible, i.e., there exists $\tilde{\mathcal{M}}$ with continuous metric such that $\mathcal{M} \subset \tilde{\mathcal{M}}$, then $\partial \mathcal{M}$, the boundary of $\mathcal{M}$ in $\tilde{\mathcal{M}}$, is a null hypersurface called a Cauchy horizon (see Fig.~\ref{fig1}). Cauchy horizons are generic in stationary black hole solutions. Furthermore, contrary to old lore, it is expected that they persist under small classical perturbations~\cite{Dafermos:2003wr, Dafermos:2012np, Dafermos:2017dbw}. For example, it is known that in Einstein-Maxwell-real scalar theory a black hole formed by collapse which asymptotes appropriately to the Reissner-Nordström (RN) metric in the exterior has a Cauchy horizon in a neighborhood of timelike infinity in the conformal completion~\cite{Dafermos:2003wr}. An analogous statement is shown for a vacuum solution which asymptotes to the Kerr metric in their exterior~\cite{Dafermos:2017dbw}. Furthermore, in maximally extended RN or Kerr black holes, the \emph{entire} Cauchy horizon persists under small perturbations~\cite{Dafermos:2012np, Dafermos:2017dbw}.\footnote{As discussed in~\cite{Dafermos:2012np, Dafermos:2017dbw}, the ``perturbed'' Cauchy horizon in all of the mentioned examples may have ``weak null singularities'' classically, which are characterized by the Christoffel symbols not being square-integrable in the continuously extended metric. The quantum singularities, discussed later in the introduction, are expected  to dominate over classical weak null singularities in many cases (see e.g.~\cite{Zilberman:2022aum}).} See Fig.~\ref{fig1}. Interestingly, quantum gravity effects can become extremely important near Cauchy horizons.

The first quantum correction to the black hole interior is obtained by analyzing quantum field theory (QFT) propagating on it as a fixed background. Since $\partial \mathcal{M}$ is not part of $\mathcal{M}$, the divergence of QFT observables (e.g., expectation values of local operators) on $\partial \mathcal{M}$ is not ruled out and, as we will discuss in detail, have in fact been derived.\footnote{To make sure that the divergence in not a coordinate artifact, one must always use coordinates that have smooth behavior at the Cauchy horizon.} A particularly significant divergence to investigate is that of $\langle T_{\mu\nu} (x)\rangle$, the expectation value of the stress-energy tensor, as $x \to \partial \mathcal{M}$, since it partly quantifies the breakdown of the semiclassical expansion. In this way, questions about the black hole singularity may be cast in terms of a QFT problem on a horizon! This offers an appealing version of the spacetime singularity problem, one which is plausibly more conducive to progress. In particular, it would be interesting to distill universal features of Cauchy horizon singularities, i.e., features that are not specific to the exact spacetime or QFT under consideration. This would pave the way for seeking out imprints of this singularity in a full quantum gravity theory, for example through the AdS/CFT correspondence~\cite{Maldacena:1997re, Witten:1998qj} (see e.g.~\cite{Fidkowski:2003nf, Festuccia:2005pi, Dodelson:2023vrw} for similar attempts in Schwarzschild-type singularities). In fact, along these lines, a breakdown of semiclassical physics near Cauchy horizons is predicted from a quantum singularity theorem proposed in~\cite{Bousso:2022tdb} which only involves universal information-theoretic conditions.

Killing Cauchy horizons (henceforth denoted by $\mathcal{H}_C$), ones whose null generators are integral curves of a Killing vector field $k^\mu$ are of particular importance given their abundance in stationary black hole spacetimes.\footnote{A stationary black hole spacetime is one which has an asymptotically timelike Killing vector field. This vector field is not equal to $k^\mu$ in general, e.g., in rotating black holes.} All charged and/or rotating black hole solutions in the Kerr-Newman family, and their asymptotically Anti-de Sitter and de Sitter counterparts give rise to Killing Cauchy horizons.\footnote{However, we do not know of a general proof that all Cauchy horizons of stationary black holes must be Killing Cauchy horizon.} In a neighborhood of $\mathcal{H}_C$, we can find smooth null coordinates $U$ and $V$, both negative in the black hole interior, and such that
\begin{align}
\mathcal{H}_C &= \{U=0,V\leq0\} \cup \{V=0,U\leq0\},\\
k^\mu &= V\partial_V - U \partial_U.
\end{align}
See Fig.~\ref{fig1}. As a black hole settles down to a stationary state, it is not only the geometry which becomes symmetric under the flow of $k^{\mu}$, but also the state of the quantum fields. We define a $k^\mu$-\emph{symmetric} state, as any state symmetric under the flow of $k^\mu$.\footnote{It is also natural to call these states stationary in the black hole exterior. However, in much of this work we discuss the black hole interior region exclusively in which $k^\mu$ is a spacelike Killing vector and the term stationary might appear misleading.} The Hartle-Hawking (HH) and Unruh states are examples of $k^\mu$-symmetric states (See e.g.~\cite{Candelas:1980zt}).\footnote{The HH state is defined as an equilibrium state (also known as KMS state~\cite{doi:10.1143/JPSJ.12.570, PhysRev.115.1342, Haag:1996hvx}) in black holes with an everywhere timelike Killing vector field in their exterior.}

Here, we adopt the philosophy that subject to a set of principles, any consistent QFT singularities at $\mathcal{H}_C$ generically happens. The observation that $\langle T_{\mu\nu} \rangle$ diverges at $\mathcal{H}_C$ was first made in the pioneering work of~\cite{BDinner}, in a toy model with spacetime dimension $d=2$ (reviewed in Sec.~\ref{sec:2}). In particular, they find
\begin{align}\label{eq-v2}
\langle T_{VV} \rangle \stackrel{V\to0^-}{=} O\left(\frac{1}{V^2}\right),
\end{align}
in the HH state. Since $T_{\mu\nu}$ is an operator of dimension $d$, in $d=2$ this singularity is the leading one compatible with dimensional analysis and the $k^\mu$-symmetry of the state which requires
\begin{align}
\langle T_{VV}( \alpha^{-1} U,V \alpha) \rangle = \alpha^2 \langle T_{VV}(U,V)\rangle,
\end{align}
for $\alpha >0$. This is consistent with our philosophy, given the absolute minimum set of principles: $k^\mu$-symmetry (of HH) and dimensional analysis.\footnote{An implicit physical assumption here is that the singularity cannot scale positively with relevant length scales of the spacetime in a region where a flat space limit is possible. This is in particular true in a neighborhood of a Killing Cauchy horizon. This is because we expect sending the length scales of the spacetime to infinity ought to make the quantum effects in the HH or Unruh states smaller. Therefore, for example, a behavior like $T_{VV} \sim \frac{r_-^2}{UV^3}$ which has the correct dimensions and is consistent with Killing symmetry is ruled out on this physical ground. The leading allowed possibility is then indeed $T_{VV} \sim V^{-2}$.}

Since the work of~\cite{BDinner}, several numerical analyses of a free scalar QFT in $k^\mu$-symmetric states on the $d=4$ RN~\cite{Lanir:2017oia, Lanir:2018vgb, Zilberman:2019buh}, Kerr~\cite{Zilberman:2022iij, Zilberman:2022aum, Zilberman:2024jns, Klein:2024sdd}, de Sitter-RN~\cite{Hollands:2019whz, Hollands:2020qpe}, and de Sitter-Kerr~\cite{Klein:2024sdd} geometries have also confirmed that $\langle T_{VV} \rangle$ diverges at $\mathcal{H}_C$.\footnote{In the HH state of the BTZ spacetime ($d=3$), this singularity is absent~\cite{Dias:2019ery}, though it has been argued that it appears at higher ordered in the semiclassical expansion~\cite{Emparan:2020rnp, Kolanowski:2023hvh}.} Though, surprisingly, the degree of divergence is still given by Eq. \eqref{eq-v2}, akin to a $d=2$ problem, and significantly milder than the naive expectation from dimensional analysis.\footnote{Of course in $d=4$, Eq. \eqref{eq-v2} is suplemented by additional length scales associated to the black hole to make it consistent with dimensional analysis.} This is a failure of the outlined philosophy which, given that $T_{\mu\nu}$ is now of dimension $4$, would have predicted a leading singularity of $T_{VV} \sim U V^{-3}$. An analogous ``mildness'' is also seen in other observables. For example, it was shown that $\langle \phi^2 \rangle$, the expectation value of the normal-ordered product of two fundamental fields, is \emph{finite} at the Cauchy horizon of RN (in Unruh and HH states)~\cite{Zilberman:2019buh} and Kerr (in the Unruh state)~\cite{Zilberman:2024jns}, and furthermore in de Sitter-Kerr (in the Unruh state)~\cite{Klein:2024sdd}:
\begin{align}\label{eq-n99n}
\langle T_{VA}\rangle \stackrel{V\to0^-}{=} O\left(\frac{1}{V}\right).
\end{align}
where $A$ denotes the azimuthally symmetric transverse direction.\footnote{See~\cite{Sela:2018xko} where the absence of only the \emph{leading} divergence in some $1$-point functions was derived by a free field WKB approximation in RN, lending more credence to the RN numerical results.}

The functional forms of these singularities are not predictable by the aforementioned philosophy. To highlight that this approach \emph{is} predictive in $d>2$ in other settings, we will discuss in Sec.~\ref{sec:2} a family of $k^\mu$-symmetric states (to which the Boulware state belongs to) where $\langle T_{\mu\nu} \rangle$ and $\langle \phi^2 \rangle$ diverge on \emph{outer} horizons in a manner completely predictable from symmetry and dimensional analysis, e.g., $\langle T_{VV}\rangle \sim U V^{-3}$  and $\langle \phi^2 \rangle \sim U^{-1} V^{-1}$ in $d=4$. This signals that when it comes to the Cauchy horizon singularities, we are missing an essential principle. The numerical results are unsatisfactory at explaining this principle, and also fail to determine what the singularity structure is for a general QFT on a general black hole Cauchy horizon in the HH or Unruh sectors. We will refer to these sets of questions as the \emph{mildness puzzle}.

\begin{figure}[htbp]
\centering
\includegraphics[width=0.55\textwidth]{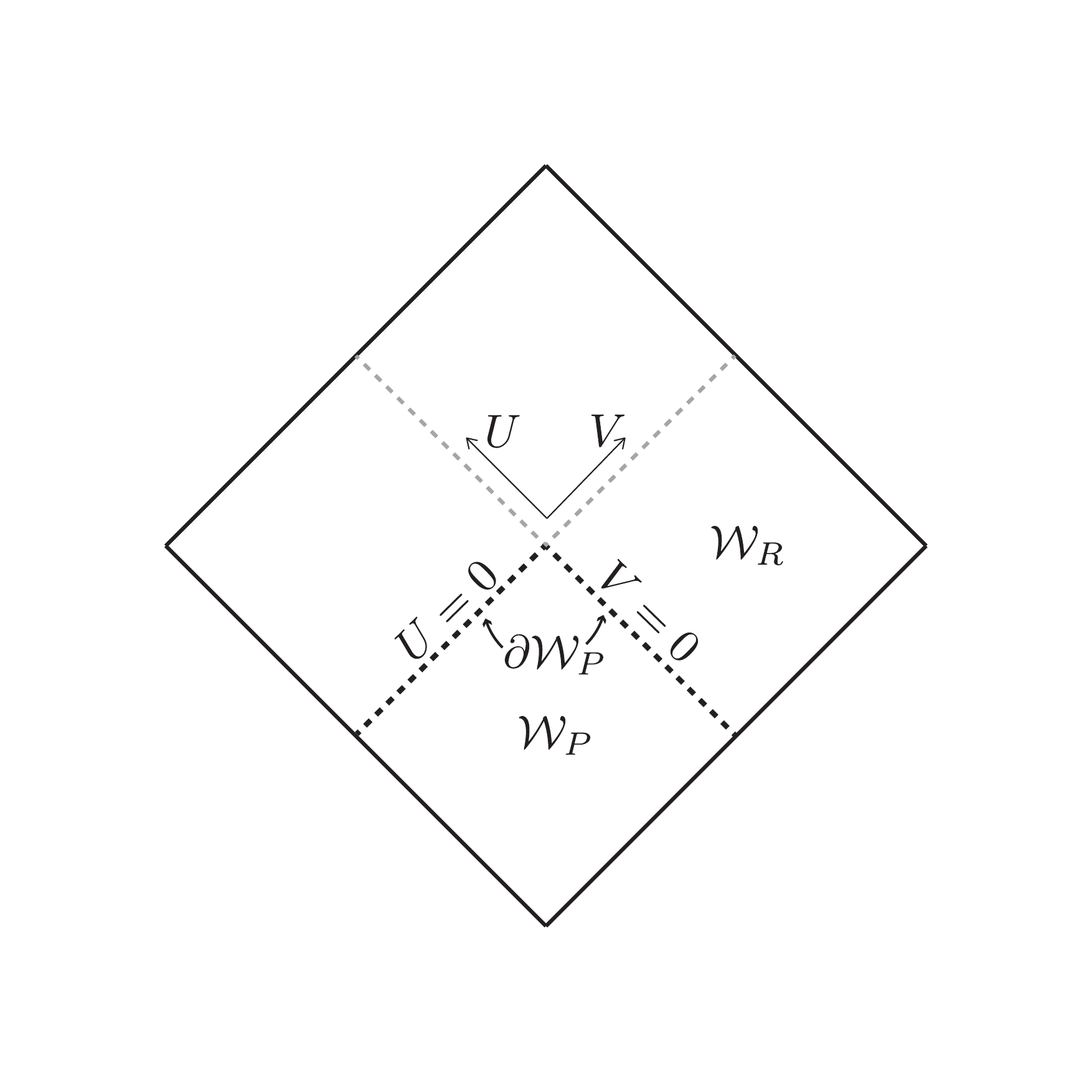} 
\caption{Minkowski space $\mathbb{R}^{1,d-1}$ is shown with the transverse directions suppressed. The region $U<0$, $V<0$ is the past Rindler wedge, denoted by $\mathcal{W}_P$, which is a simple example of a spacetime with a Killing Cauchy horizon $\partial \mathcal{W}_P$ shown by the black dotted lines. The region $U<0$, $V>0$ is the more familiar right Rindler wedge, denoted by $\mathcal{W}_R$. }
\label{fig2}
\end{figure}

In this paper, with our sights set on explaining the mildness puzzle, we develop general QFT tools to analyze Cauchy horizon singularities in $d>2$.\footnote{The reason for this restriction is that this is where the mildness puzzle only manifests for $d>2$.} Our main focus is on (arguably) the simplest spacetime with a complete Cauchy slice which gives rise to a Cauchy horizon: the Past Rindler Wedge. Consider the Minkowski spacetime coordinates $(U,V,y^A)$ in which the metric is
\begin{align}\label{eq-2c124523}
ds^2 = -dU dV + \sum_{A=1}^{d-2} (dy^A)^2.
\end{align}
The past Rindler wedge, henceforth denoted by $\mathcal{W}_P$, is the following region (see Fig.~\ref{fig2}):
\begin{align}\label{eq-nv249j2}
\mathcal{W}_P = \{(U,V,y^A)| U<0, V<0\}.
\end{align}
Note that $k^\mu = V\partial_V-U \partial_U$ is a Killing vector field in $\mathcal{W}_P$.\footnote{To our knowledge, the term ``past Rindler wedge'' is new. It should not be confused with the better-known Milne wedge which is the past of a point in flat spacetime.} The Killing Cauchy horizon of $\mathcal{W}_P$ is $\partial \mathcal{W}_P$, i.e., the boundary of $\mathcal{W}_P$ as a subset of $\mathbb{R}^{1,d-1}$. Besides being a simple example of a spacetime with a Cauchy horizon, $\mathcal{W}_P$ is also the ``near $\mathcal{H}_C$'' limit of \emph{any} black hole spacetime with an $\mathcal{H}_C$.\footnote{A black hole spacetime with a Cauchy horizon is not necessarily extendible to a larger globally hyperbolic spacetime, but $\mathcal{W}_P$ is extendible to Minkowski spacetime. Nevertheless, the near Cauchy horizon region of a black hole is $\mathcal{W}_P$. Furthermore, as we discuss in Sec.~\ref{sec:5}, the black hole interior region by itself (region $II$ in Fig.~\ref{fig1}) can be smoothly extended to a larger globally hyperbolic spacetime.} Therefore, any general lesson we learn about $\partial \mathcal{W}_P$ singularities automatically applies to the singularity structure of \emph{all} black hole Cauchy horizons after a strict near Cauchy horizon limit is taken (see Sec.~\ref{sec:5} for details). This motivates us to take it seriously as a model to study black hole Cauchy horizon singularities. Though $\mathcal{W}_P$ is our main focus, in Sec.~\ref{sec:5} we discuss how to apply our ideas to a black hole Cauchy horizon without the strict $\mathcal{W}_P$ limit.

\subsection*{Outline and summary of main points}

We will now very briefly summarize the paper's main points as we outline its structure. In Sec.~\ref{sec:2}, we review a simple derivation of a $d=2$ Cauchy horizon singularity and explicitly derive Eq.~\eqref{eq-v2}. We then juxtapose it with stationary black hole exterior states that are out-of-equilibrium with the black hole (e.g., the Boulware state), where the black hole \emph{outer} horizon has quantum singularities. Their singularity structure is not mild, i.e., it is perfectly predictable from symmetry and dimensional analysis. For example,
\begin{align}
\langle T_{VV}\rangle \stackrel{V\to0^-}{=} O\left(\frac{1}{UV^3}\right),
\end{align}
in $d=4$, where the $U,V$ coordinates are outer horizon Kruskal coordinates such that $V=0$ ($U=0$) is the past (future) horizon. We explain that in a Euclidean continuation of the black hole geometry, the singularity can be attributed to the existence of a conical singularity at the outer horizon. We further show that since the Cauchy horizon and the outer horizon have different surface gravities generically, even in the HH state where the outer horizon is smooth, there exists an analogous conical singularity at the Cauchy horizon, in a complexified geometry. Despite this tight analogy between outer and Cauchy horizons, the Cauchy horizon singularity structure is significantly milder, further highlighting the mildness puzzle.

We then focus exclusively on Lorentzian signature. In Sec.~\ref{sec:3}, guided by an axiomatic QFT framework (mostly following~\cite{Hollands:2009bke}), we analyze QFT singularities on the boundaries of a spacetime \emph{wedge} $\mathcal{W}$, i.e., an open region of maximal Cauchy evolution within a larger globally hyperbolic spacetime $\mathcal{W}_0$. Explicitly, after a brief review of the relevant subset of QFT axioms in Sec.~\ref{sec:31}, in Sec.~\ref{sec:32} we propose a very broad construction of QFT states with singularities on $\partial \mathcal{W}$. The construction defines a new state $\omega$ of $\mathcal{W}$, starting from a $\mathcal{W}_0$ state $\omega_0$, which is smooth on $\partial \mathcal{W}$, and creates singularities on $\partial \mathcal{W}$ by placing operator-valued distributions in the causal complement $\mathcal{W}^c$, i.e., the set of $\mathcal{W}_0$ points whose domain of influence does not intersect $\mathcal{W}$ (see Fig.~\ref{fig3}). Explicitly, we have
\begin{align}\label{eq-24v2b6}
\langle \Phi^{(i_1)}(x_1) \cdots \Phi^{(i_n)}(x_n) \rangle^{\mathcal{W}}_\omega = \langle \mathcal{D}(\mathcal{W}^c) \Phi^{(i_1)}(x_1) \cdots \Phi^{(i_n)}(x_n)\rangle^{\mathcal{W}_0}_{\omega_0},
\end{align}
where $\Phi^{(i)}(x)$ are abstractly labelled local operators of a general QFT and $\mathcal{D}(\mathcal{W}^c)$ denotes the singularity-generating operator (see Sec.~\ref{sec:31} for details). The relation \eqref{eq-24v2b6} constructs a new set of $\mathcal{W}$ $n$-point functions from existing ones in the $\omega_0$ state of $\mathcal{W}_0$. We also present a generalization of Eq. \eqref{eq-24v2b6} applicable when $\mathcal{W}^c = \emptyset$, but $\partial \mathcal{W} \neq \emptyset$.\footnote{As would be the case for $\mathcal{W}_P$ as a subset of Minkowski spacetime.}

That the singularity-generating operators \emph{must} be contained in $\mathcal{W}^c$ comes from ensuring consistency with the microlocal spectrum condition in QFT (reviewed in Appendix~\ref{sec:app1}) which constrains the structure of $n$-point function singularities. This condition plays the key role in our analysis to distinguish the physics of the outer horizon from the Cauchy horizon. To test the validity of our construction, in Sec.~\ref{sec:33} we provide various examples of it for a free massless scalar field, where we can explicitly check that the QFT axioms are satisfied.

The construction~\eqref{eq-24v2b6} is extremely general since it allows arbitrary $\omega_0$ and arbitrary operators $\mathcal{D}(\mathcal{W}^c)$ to construct $\mathcal{W}$ states. The apparent generality of construction \eqref{eq-24v2b6} is further substantiated in subsection~\ref{sec:41} were we show that a natural class of conformal field theory state in $\mathcal{W}_P$ \emph{all} arise as special cases of construction \eqref{eq-24v2b6}. Furthermore, the out-of-equilibrium outer horizon singularities discussed in Sec.~\ref{sec:2} all arise as special cases of this construction. In light of these, we take the entire family of states constructible via Eq.~\eqref{eq-24v2b6} as our model for black hole Cauchy horizon singularities. Remarkably, we find that these states satisfy a mild singularity structure on $\partial \mathcal{W}_P$, analogous to the aforementioned numerical results. We view this as strong evidence that indeed black hole Cauchy horizon states, like outer horizon singular states, are reproducible from construction \eqref{eq-24v2b6}. This may appear miraculous, though a minimal plausible explanation for this would be if all $\mathcal{W}_P$ states are indeed constructible via Eq. \eqref{eq-24v2b6}. Though we do not wish to make this claim without further investigation.

Let us outline the rough argument for the derivation of the mild singularity ansatz. Construction \eqref{eq-24v2b6} applied to $\mathcal{W}_P$ in particular says:\footnote{As we will explain in more detail in Sec.~\ref{sec:32}, to apply construction Eq. \eqref{eq-24v2b6} to $\mathcal{W}_P$, we need to slightly generalize it to allow limits of operators taken to the asymptotic boundary. We will not expand on that detail in this brief overview in the introduction.}
\begin{align}
\langle \Phi^{(i)}(x) \rangle^{\mathcal{W}_P}_\omega = \sum_{m_1 m_2} c_{m_1 m_2} \langle \Gamma^{(m_1)} (\mathcal{W}^+_\infty)\Gamma^{(m_2)}(\mathcal{W}^-_\infty)  \Phi(x)\rangle^{\mathbb{R}^{1,d-1}}_{\omega_0}
\end{align}
Here, $c_{m_1 m_2}$ are complex numbers and $\Gamma^{(m_1)}(\mathcal{W}^+_\infty)$ and $\Gamma^{(m_2)}(\mathcal{W}^-_\infty)$ are the limits of singularity-generating operators whose support in this limit goes to $\mathcal{W}^+_\infty$ and $\mathcal{W}^-_\infty$ (see Sec.~\ref{sec:32} for detailed definition). $\mathcal{W}^+_\infty$ and $\mathcal{W}^-_\infty$ are the two connected components of the largest subset of points on past null infinity of Minkowski spacetime whose causal future does not intersect $\mathcal{W}_P$ (see Fig.~\ref{fig5}). Importantly, $\mathcal{W}^+_\infty$ ($\mathcal{W}^-_\infty$) is spacelike-separated from the region $V<0$ ($U<0$). As $\Phi^{(i)}(x)$ approaches $V=0$ from within $\mathcal{W}_P$, suppose it satisfies:
\begin{align}\label{eq-2v4b2}
\langle \Phi^{(i)}(U,V,y^A) \rangle^{\mathcal{W}_P}_{\omega} \stackrel{V\to 0^-}{=} \sum_{m_1 m_2} c_{m_1 m_2} \langle \Gamma^{(m_2)}(\mathcal{W}^-_\infty)\rangle^{\mathbb{R}^{1,d-1}}_{\omega_0} \langle \Gamma^{(m_1)} (\mathcal{W}^+_\infty)  \Phi^{(i)}(x)\rangle^{\mathbb{R}^{1,d-1}}_{\omega_0} + \cdots,
\end{align}
where ellipsis denotes subleading terms in a small $V$ expansion. The singularity structure dictated by Eq.~\eqref{eq-2v4b2} is mild because in particular the leading term on the RHS has a smooth extension to the $U=0$ horizon. This is because $\Gamma^{(m_1)} (\mathcal{W}^+_\infty)$, being spacelike separated from the $U=0$ portion of $\partial \mathcal{W}_P$, cannot cause singularities there. This smooth extension to $U=0$ is for example satisfied by the aforementioned Cauchy horizon singularity $T_{VV}\sim V^{-2}$, but not by a stronger singularity like $T_{VV}\sim U V^{-3}$. In Sec.~\ref{sec:42}, we define a notion of a robust singularity, i.e., a $1$-point function singularity at the Cauchy horizon whose leading term in the asymptotic expansion as $V\to0^-$ is state-independent. By extending, as an assumption, an argument involving the operator product expansion of local operators to the more general operators $\Gamma^{(m)}(\mathcal{W}^{\pm}_\infty)$, we demonstrate that the robustness of the singularity would in particular imply the factorized structure \eqref{eq-2v4b2}. We discuss additional criteria, other than robustness, which also leads to Eq.~\eqref{eq-2v4b2}.

Finally, combined with $k^\mu$-symmetry, we derive the following singularity structure (see Sec.~\ref{sec:4} for details): For any operator $\Phi_{\mu_1, \cdots, \mu_r}(x)$, with the indicated tensor structure,\footnote{Here, we deviate from the abstract notation $\Phi^{(i)}$, and use the more familiar notation where the tensor structure of an operator is explicit.} the \emph{allowed} robust singularity as $V \to 0^-$ satisfies:\footnote{Note that our analysis does not force the allowed singularities to necessarily appear. It merely forces more sever singularities to not appear.}
\begin{align}
\langle \Phi_{\mu_1, \cdots, \mu_r}(U,V,y^A) \rangle^{\mathcal{W}_P}_{\omega} &\stackrel{V\to0^-}{=}O\left(\frac{1}{V^k}\right), \quad  \text{for  } k>0,\label{eq-mainmain11111} \\ 
\langle \Phi_{\mu_1, \cdots, \mu_r}(U,V,y^A) \rangle^{\mathcal{W}_P}_{\omega} &\stackrel{V\to0^-}{=}o\left(\frac{1}{V^\epsilon}\right), \quad  \text{for  } k\leq0.\label{eq-mainmain22222}
\end{align}
where $\epsilon$ is any positive number and $k$ is the number of $V$ indices minus the number of $U$ indices in the specific component of $\Phi_{\mu_1, \cdots, \mu_r}(x)$ under consideration. The singularity structures Eq.~\eqref{eq-v2} and \eqref{eq-n99n} are obvious special cases for $r=2$ and $r=1$, as is the aforementioned absence of a power law divergence in $\phi^2$ at the Cauchy horizon.\footnote{Notably, our constraints still would in principle allow a logarithmic divergence in $\phi^2$ at the Cauchy horizon.} The ansatz applies more generally to any $\Phi_{\mu_1, \cdots, \mu_r}(x)$ which satisfies Eq.~\eqref{eq-2v4b2}, whether it has a robust singularity or not.

A prediction of our analysis for all black hole Cauchy horizons in the HH or Unruh state is that any robust Cauchy horizon singularity that survives the strict $\mathcal{W}_P$ limit, must fall under the ansatz \eqref{eq-mainmain11111} and \eqref{eq-mainmain22222}. We will expand on the details in Sec.~\ref{sec:5}, but a particular upshot is that a $d>2$ black hole Cauchy horizon (in the HH or Unruh sectors) is forbidden to have a robust singularity of the form:
\begin{align}
\langle T_{\mu\nu} (U,V, y^A)\rangle =O\left(\frac{1}{U^{\frac{d-k}{2}} V^{\frac{d+k}{2}}}\right)
\end{align}
where $k$ is the number of $V$ indices minus the number of $U$ indices in the stress tensor component under consideration. This is fully consistent with existing numerical results that we are aware, but predicts non-trivial constraints for the infinite set of spacetimes and QFTs whose Cauchy horizon singularities have not been analyzed with a direct computation yet.

Let us emphasize that our work provides only a partial answer to the mildness puzzle, since it rests on the assumption that all black hole Cauchy horizon states in the strict near Cauchy horizon limit (i.e. the $\mathcal{W}_P$ limit) are instantiations of our construction~\eqref{eq-24v2b6}. Though we view the emergence of the mild singularity ansatz as strong evidence for this. This assumption may be lifted if one could prove that construction~\eqref{eq-24v2b6} can create arbitrary $\mathcal{W}_P$ state. We made partial progress toward this goal in Sec.~\ref{sec:41} and leave further investigations to future work. Another limitation of our result is that we cannot attest a priori whether a Cauchy horizon singularity is robust or not. However, in~\cite{Hollands:2019whz, Hintz:2023pak, Klein:2024sdd} it was argued that the $\langle T_{VV} \rangle$ and $\langle T_{VA}\rangle$ singularities in the Unruh state of de-Sitter RN and de-Sitter Kerr are in fact robust by our definition. We therefore find it plausible that an independent argument for the robustness of certain black hole Cauchy horizon singularities can be made to complete the story. Alternatively, we find it plausible that the structure \eqref{eq-2v4b2} can be derived more generally and so the robustness condition can be relaxed.

Finally, in Sec.~\ref{sec:5}, we outline steps needed to apply these ideas more directly to a black hole Cauchy horizon, without the need for the strict $\mathcal{W}_P$ limit.


\section{Horizon Singularities and the Rindler Wedge Limit}
\label{sec:2}

In this section, we elaborate on some of the basic ideas presented in the introduction. We review a simple derivation of a Cauchy horizon singularity in a two-dimensional toy model, compare its degree of divergence with the $d=4$ numerical results, and contrast with a family of easy-to-analyze singular \emph{outer} horizon states. This will highlight the mildness puzzle.

In~\cite{BDinner}, the authors consider the HH state on a $d=2$ ``RN-like'' spacetime given by the metric
\begin{align}\label{eq-2dRN}
ds^2 = -f(r) dt^2 + f(r)^{-1} dr^2,
\end{align}
where
\begin{align}
f(r) = \left(1-\frac{r_+}{r}\right)\left(1-\frac{r_-}{r}\right).
\end{align}
Like RN, the $r>r_-$ portion of this geometry is globally hyperbolic and gives rise to a $\mathcal{H}_C$ at $r=r_-$.\footnote{The corresponding Penrose diagram would be as in right Fig.~\ref{fig1}.} For any stationary black hole where the exterior (e.g., region $I$ in Fig.~\ref{fig1}) has an everywhere timelike Killing vector field $\partial_t$, one can define a thermal equilibrium state (also known as the KMS state~\cite{doi:10.1143/JPSJ.12.570, PhysRev.115.1342, Haag:1996hvx}) in the exterior characterized by
\begin{align}
\langle \Phi^{(i_1)}(t_1) \Phi^{(i_2)}(t_2)  \rangle_\beta = \langle \Phi^{(i_2)}(t_2) \Phi^{(i_1)}(t_1+ i \beta)  \rangle_\beta
\end{align}
where $\Phi^{(i_1)}(t_1)$ and $\Phi^{(i_2)}(t_2)$ denote arbitrary local operators at the indicated times and arbitrary (omitted) points in space. The HH state is the special case with $\beta = 2\pi/\kappa_+$, where $\kappa_+$ is the surface gravity of the outer horizon, given by:
\begin{align}
\kappa_+ = \frac{f'(r_+)}{2}
\end{align}
This special choice of $\beta$ leads to smooth behavior of the QFT observables at the outer horizon which enables the continuation of the state to the black hole interior. The HH state can equivalently be defined by performing a Euclidean path integral on the $t \to i \tau$ continuation of geometry, with the identification $\tau \sim \tau+\beta$. The special choice $\beta = 2\pi/\kappa_+$ is the only one for which the Euclidean geometry is free of a conical singularities at $r=r_+$.

An important feature of the HH state is the vanishing of the energy flux through the past and future outer horizons. For any $d=2$ CFT with central charge $c$, the trace anomaly and conservation equations,
\begin{align}
&\langle T^\mu_\mu \rangle=\frac{c}{12} R,\\
&\nabla_\mu \langle T^{\mu\nu} \rangle=0,
\end{align}
where $R$ is the Ricci scalar, determine $\langle T_{\mu\nu} \rangle$ up to two constants, which the no-flux condition of HH fixes. Then, it is an easy exercise to obtain $\langle T_{\mu\nu} \rangle$ near $\mathcal{H}_C$~\cite{BDinner}. In particular, one finds:
\begin{align}\label{eq-exact2dT}
\langle T_{VV} \rangle\stackrel{V\to0^-}{=}\frac{c}{48 \pi^2 V^2} \left(\frac{\kappa_-^2}{\kappa_+^2}-1\right)
\end{align}
where $\kappa_- = |f'(r_-)|/2$. Here, $V$ belongs to the the inner Kruskal coordinates $(U,V)$, which approach the $d=2$ Minkowski coordinates (see Eq.~\eqref{eq-2c124523}) in a neighborhood of $r=r_-$.\footnote{The definition of the inner Kruskal $(U,V)$ is as follows: Let $dr^* = \int \frac{dr}{|f(r)|}$, from which we define $v = t+r^*$, $u = -t+r^*$. Then, $V \sim -\exp(-\kappa_- v)$ and $U \sim -\exp(\kappa_- u)$, where $\sim$ means equality up to an arbitrary multiplicative dimensionful constant.} The expression inside the parenthesis in Eq. \eqref{eq-exact2dT} is reminiscent of factors which appear in other well-known $\langle T_{\mu\nu} \rangle$ divergences near \emph{outer} horizons. They happen if the exterior of the horizon is in a KMS state with the ``wrong temperature''. For example, consider the exterior of a black hole, in arbitrary $d$, in a KMS state with $\beta \neq 2\pi/\kappa_+$. In the Euclidean preparation of this state, this choice of $\beta$ results in the presence of a conical singularity on the horizon. This is a place where QFT observables can diverge. In fact, this is an effect which can be analyzed purely in the near-horizon limit where the geometry approaches that of a Rindler wedge of flat spacetime, denoted by $\mathcal{W}_R$ (see Fig.~\ref{fig2}):
\begin{align}
\mathcal{W}_R =\{(U,V,y^A)|V>0, U<0 \}
\end{align}
The ``wrong temperature'' KMS state above in the near-horizon limit approaches the KMS state of $\mathcal{W}_R$ with respect to $\partial_t= V \partial_V-U\partial_U$, with $\beta = 2\pi/z$ and $z \neq 1$.\footnote{The case with $z=1$, is the ``right temperature'' and corresponds to the Minkowski vacuum restricted to $\mathcal{W}_R$.} Let us write down this state of $\mathcal{W}_R$ for a free massless scalar quantum field $\phi$. This is indeed a Gaussian state where the two-point function can be calculated analytically for all $z\geq0$, though for integer $z$ it is easily derivable using the method of images~\cite{Calabrese:2004eu}. For example, in $d >2$ for $z=2$:
\begin{align}\label{eq-13e1f4}
\langle \phi(x_1) \phi(x_2) \rangle &=G_0((U_1,V_1,y^A_1),(U_2,V_2,y^A_2))+G_0((U_1,V_1,y^A_1),(-U_2,-V_2,y^A_2)),
\end{align}
where the points $x_1,x_2 \in \mathcal{W}_R$ are specified by their $(U,V,y^A)$ coordinates, and
\begin{align}\label{eq-2rf124}
G_0(x_1,x_2)= \frac{\Gamma(\frac{d}{2}-1)}{4 \pi^{\frac{d}{2}}[-(U_1-U_2)(V_1-V_2)+ \sum_{A=1}^{d-2} (y_1^A-y_2^A)^2 + i \epsilon (U_2+V_2 - U_1-V_1)]^{\frac{d-2}{2}}}.
\end{align}
where the numerator is the Gamma function. In Eq.~\eqref{eq-13e1f4}, the ``image'' is located in the region $U>0$, $V<0$. Note that $G_0(x_1,x_2)$ is the Minkowski vacuum two-point function and is defined as a distribution via the $\epsilon \to 0$ limit. From Eq. \eqref{eq-13e1f4} (or its general $z$ extension) it is easy to compute $1$-point functions like $\langle T_{VV} \rangle$. The formula is especially simple for the conformally coupled theory:\footnote{For a minimally coupled scalar field, we have $T_{VV} = (\partial_V\phi)^2$, while for a conformally coupled theory we have $T_{VV} = (\partial_V\phi)^2 - \frac{d-2}{4(d-1)} \partial_V^2 \phi^2$, where $\phi^2$ is the normal-ordered product.}
\begin{align}\label{eq-WRsing}
\langle T_{VV} \rangle = a\frac{z^d-1}{(U V)^{\frac{d-2}{2}} V^2}.
\end{align}
where $a$ is an arbitrary $z$ independent real number.

Let us extract some lessons from this detour into the ``wrong temperature'' outer horizon singularities. In light of the factor of $z^d-1$ in Eq. \eqref{eq-WRsing}, whose $d=2$ version appeared in the $\mathcal{H}_C$ singularity \eqref{eq-exact2dT}, it is tempting to speculate that the $\mathcal{H}_C$ singularities in general can be thought of as the Cauchy horizon being in a state of ``wrong temperature'' (see also~\cite{Balasubramanian:2019qwk, Papadodimas:2019msp, Chen:2024ojv} for arguments along these lines). However, the functional form of the divergence in \eqref{eq-WRsing} is commensurate with and in fact fully predictable from dimensional analysis and $k^\mu$-symmetry. In particular, its divergence gets more severe as $d$ increases, in sharp contrast with the previously mentioned free field numerical analysis of Cauchy horizon singularities in $d=4$~\cite{Lanir:2017oia, Lanir:2018vgb, Zilberman:2019buh,Zilberman:2022iij, Zilberman:2022aum, Zilberman:2024jns, Klein:2024sdd,Hollands:2019whz, Hollands:2020qpe}.

In fact, we can push the analogy between Cauchy horizons and the ``wrong temperature'' states even further. Consider the Euclidean continuation of the geometry~\eqref{eq-2dRN}, corresponding to the HH state:
\begin{align}
ds^2 = f(r)d\tau^2 + f(r)^{-1} dr^2
\end{align}
for $r>r_+$ where $\tau \sim \tau+\beta$. One can continue this geometry from the $r \geq r_+$ region into the $r_-\leq r<r_+$ region while maintaining periodicity $\tau \sim \tau+\beta$.\footnote{To our knowledge, this type of continuation was first discussed in~\cite{Candelas:1985ip}, but for the case of a Schwarzschild black hole. Also, to get a connected geometry one needs to complexify $r$ in a small neighborhood around $r=r_+$, since otherwise the geometry would cap off at $r=r_+$ and not connect to the interior portion.} This new piece of the geometry has a $(-,-)$ signature. Interestingly, one finds in this geometry a conical singularity at $r=r_-$ analogous to the one which appear at $r=r_+$ in the ``wrong temperature'' exterior states. This behavior trivially generalizes to $d>2$. The appearance of this conical singularity naively suggests that the quantum singularities at the Cauchy horizons are similar physically to the outer horizon in the aforementioned wrong temperature states. Though, this analogy falls apart once we compare the degree of the divergences at the horizon, again highlighting the mildnesss puzzle.\footnote{Importantly, the geometry in which the Cauchy horizon has a conical singularity violates the Kontsevich-Segal-Witten criterion~\cite{Kontsevich:2021dmb, Witten:2021nzp} which could explain the physical difference between the Cauchy horizon and the outer horizon.}

Similar to the outer horizon analysis above, it is useful to think about the singularity structure of $\mathcal{H_C}$, by taking a ``near $\mathcal{H_C}$'' limit. The key difference is that this limit results in the past Rindler wedge $\mathcal{W}_P$, defined in Eq. \eqref{eq-nv249j2}. Despite the superficial similarity between $\mathcal{W}_R$ and $\mathcal{W}_P$, they are physically quite different, especially in $d>2$, where the Cauchy horizon mildness puzzle arises. Let us discuss some intuitive physical differences between the wedges. Recall the notion of the causal complement of a spacetime region, i.e., the set of points spacelike separated from it. While the causal complement of $\mathcal{W}_R$ is the infinitely large region $(U\geq0,V\leq0)$, the causal complement of $\mathcal{W}_P$ is the empty set! Roughly speaking, $\mathcal{W}_P$ knows more of Minkowski spacetime than $\mathcal{W}_R$ does.\footnote{In the Haag-Kastler axiomatic framework of QFT, Haag duality roughly states that the in the vacuum sector of Minkowski spacetime, the algebra of observables associated to spacetime subregion is the same as that of the causal complement of its causal complement (Eq.~(III.4.9) in tentative postulate 4.2.1 in~\cite{Haag:1996hvx}). Assuming this, it follows from the emptiness of the causal complement of $\mathcal{W}_P$, that the algebra of observables restricted to $\mathcal{W}_P$ is equivalent to that of the infinitely larger Minkowski spacetime. This might tempt the reader to conclude that no $\mathcal{W}_P$ state can have $\partial \mathcal{W}_P$ singularities. However, in Sec.~\ref{sec:33}, we will explicitly construct QFT states with singularities on $\partial \mathcal{W}_P$, which consequently are not in the vacuum sector of Minkowski spacetime.}

Another heuristic difference arises if one naively tries to define the analogue of the $\mathcal{W}_R$ $2$-point function \eqref{eq-13e1f4} in $\mathcal{W}_P$. Then, the corresponding ``image'' would have to be placed in the $V>0$, $U>0$ region (see Fig.~\ref{fig2}). In $d>2$ this leads to ``in-wedge'' singularities in the two-point function of the fundamental field as, say $x_2$ is held fixed and $x_1$ is taken to a hit the past light cone of the image of $x_2$. This singularity can arise even when $x_1$ and $x_2$ are spacelike separated from each other, which is not allowed by the QFT axioms. See Sec.~\ref{sec:31} for details. Therefore, there is no direct analogue of the wrong temperature KMS state in $\mathcal{W}_P$.

In light of these observations, we expect that the allowed QFT singularities on the boundaries of wedges can sensitively depend on the geometry of the wedge. Even though $\mathcal{W}_P$ is the main spacetime of interest in this paper, we find it fruitful to analyze QFT in a more general wedge which includes $\mathcal{W}_P$ (and $\mathcal{W}_R$) as a special case in order to better tease out the general principles behind allowed QFT singularities. This will be the main focus of the next section.

\section{A Proposal for the Singularity Structure on the Boundaries of Wedges}
\label{sec:3}

Here, we will first review the basic principles of QFT inside a smooth globally hyperbolic spacetime $\mathcal{W}$ in subsection~\ref{sec:31}. Then, we will restrict to $\mathcal{W}$ which is an open subset of a larger globally hyperbolic spacetime $\mathcal{W}_0$ in which it is maximally Cauchy evolved. This setup obviously includes $\mathcal{W}_R$ and $\mathcal{W}_P$ as special cases of $\mathcal{W}$, with $\mathcal{W}_0 = \mathbb{R}^{1,d-1}$. In Sec.~\ref{sec:32}, guided by the QFT axioms, we propose a broad construction of $\mathcal{W}$ states with $\partial \mathcal{W}$ singularities, which descend from $\mathcal{W}_0$ states. Various examples of this construction are then provided in Sec.~\ref{sec:33}. In Sec.~\ref{sec:4}, we show that this construction applied to $\mathcal{W}_P$ leads to a singularity structure analogous to that of a black hole Cauchy horizon interior.

\subsection{Quantum field theory in a wedge}
\label{sec:31}

First, let us review the main consistency conditions a QFT operator algebra and its states needs to satisfy. Since our analysis is not limited to a specific QFT, we find it useful to keep things general and follow an axiomatic Wightman-type approach to analyzing a QFT in $\mathcal{W}$, mostly following~\cite{Hollands:2009bke}, to which we refer the reader for extra details. Readers familiar with these basics may skip to the next subsection.

To define a QFT in $\mathcal{W}$ (with $d>2$)\footnote{We focus on $d>2$ since this is where the mildness puzzle arises.}, we demand a collection of local operators $\Phi^{(i)}(x)$, for $x \in \mathcal{W}$, where the label $i$ belongs to a discrete set $I$ (this includes the identity operator $\mathbb{1}$).\footnote{More precisely, $\Phi^{(i)}(x)$ are operator-valued distributions} Moreover, there is a map $\star: I \to I$, preserving $\mathbb{1}$, and such that for each $\Phi^{(i)}(x)$, there exists a starred counterpart ${\Phi^{(i)}}^*(x)$ and further the $*$ map squares to identity, i.e., ${\Phi^{(i)}}^{**}(x)=\Phi^{(i)}(x)$. The set of local fields include fundamental fields, composite fields, and their derivatives. So, for example, in a free scalar QFT, $\Phi^{(i)}$ could denote the fundamental field $\phi$ itself, or it could denote $\phi^2$, $\nabla_\mu \phi$, $T_{\mu\nu}$ for specific choices of $\mu$, $\nu$, and so on. This notation leaves implicit any spinor or tensor structure that an operator $\Phi^{(i)}(x)$ may be a components of.\footnote{We need also to demand a spin structure on $\mathcal{W}$ if fermions are present.}

A main ingredient is the set of all $n$-\emph{point functions} in a state $\omega$ in $\mathcal{W}$, denoted by
\begin{align}\label{eq-npoint}
\langle \Phi^{(i_1)}(x_1) \cdots \Phi^{(i_n)}(x_n) \rangle^{\mathcal{W}}_\omega,
\end{align}
which is unit normalized, i.e., $\langle \mathbb{1} \rangle_\omega=1$. The $n$-point functions \eqref{eq-npoint} are in fact distributions $E : V(i_1) \times \cdots \times V(i_n) \to \mathbb{C}$ where $V(i)$ are the bundles over $\mathcal{W}$ which correspond to the tensor/spin structure of $\Phi^{(i)}$. Integrating these $n$-point distributions against suitable test (or smearing) functions $f(x)$ yields complex numbers. Therefore, the set \eqref{eq-npoint} for all (non-negative) integers $n$ yields a linear map $\omega: \mathcal{A} \to \mathbb{C}$, where $\mathcal{A}$ denotes an operator algebra generated by the set of all smeared fields
\begin{align}
\Phi^{(i)}(f) =\int dx~f(x) \Phi^{(i)}(x).
\end{align}
for all $\Phi^{(i)}$ and suitable smearing functions $f(x)$. This integral representation of elements $\Phi^{(i)}(f)$ is somewhat formal as it makes sense inside correlation functions, but it makes manifest the linearity property $\Phi^{(i)}(a f_1+b f_2) = a\Phi^{(i)}(f_1) + b\Phi^{(i)}(f_2)$, where $a,b \in \mathbb{C}$. This algebra $\mathcal{A}$ is in fact a $*$-algebra in which the $*$ operation is defined by
\begin{align}
\Phi^{(i)}(f)^* = {\Phi^{(i)}}^*(\bar{f}),
\end{align}
where $\bar{f}$ denote the complex conjugate of $f$.

A fundamental constraint on QFT data (which is indirectly a constraint on the $n$-point functions) is the positivity of the state, i.e.,
\begin{align}\label{eq-23cir15}
\langle a^* a \rangle^{\mathcal{W}}_\omega \geq 0, \quad & \forall a \in \mathcal{A}.
\end{align}
The state $\omega$ is then a positive linear map on the QFT algebra from which it is possible to construct the so-called GNS Hilbert space $\mathcal{H}_\omega$ with a representation, $\pi :\mathcal{A} \to \pi(\mathcal{A})$, in which for $a \in \mathcal{A}$, the Hermitian adjoint is given by $\pi(a)^\dagger = \pi(a^*)$ and such that there exists $\ket{\Omega} \in \mathcal{H}_\omega$ satisfying
\begin{align}
\langle a \rangle^{\mathcal{W}}_\omega = \bra{\Omega} \pi(a) \ket{\Omega}, \quad & \forall a \in \mathcal{A}.
\end{align}
Acting with $\pi(a)$ on $\ket{\Omega}$ for all $a\in \mathcal{A}$ spans a dense subspace of $\mathcal{H}$~\cite{Haag:1996hvx}.

Besides positivity, a significant constraint on the $n$-point functions, which will be our primary focus here, pertains to their allowed ``in-wedge'' singularities (not to be confused with singularities as operators approach $\partial \mathcal{W}$). For example, if $\mathcal{W}$ is smooth, $1$-point functions are constrained to be smooth functions.\footnote{These smooth functions are allowed to diverge in the limit of points ``leaving'' the spacetime such at the boundary of the wedge.} However, generally speaking, for $n\geq 2$ singularities can appear in the $n$-point functions, for example, as two of the operators coincide or become light-like separated. It turns out that placing natural local positivity conditions on the energy spectrum constrains the structure of these singularities. These constraints have been framed in terms of properties of the Fourier transform of the $n$-point functions, and are known as the microlocal spectrum conditions (MSC)~\cite{Radzikowski:1996pa, Brunetti:1995rf, Brunetti:1999jn, Strohmaier:2002mm}. We will review the MSC in Appendix~\ref{sec:app1} for completeness, and in what follows simply state the only feature of MSC that is of importance to us.

Lastly, in the axiomatic framework of \cite{Hollands:2009bke}, the operator product expansion (OPE) plays a crucial role. The OPE says:
\begin{align}\label{eq-OPE}
\langle \Phi^{(i_1)}(x_1) \cdots \Phi^{(i_n)}(x_n) \rangle^{\mathcal{W}}_\omega = \sum_{j}C^{(i_1)\cdots (i_n)}_{(j)}(x_1,\cdots,x_n;y) \langle \Phi^{(j)}(y) \rangle^{\mathcal{W}}_\omega
\end{align}
as $x_1, \cdots, x_n \to y$. The OPE coefficients $C^{(i_1)\cdots (i_n)}_{(j)}$ can diverge in this limit and the equality is only meant in an asymptotic expansion sense~\cite{Hollands:2009bke}. In this section, we do not need a detailed analysis of the OPE, though it will play an important role in Sec.~\ref{sec:4}. However, the general principle behind the construction of the next subsection is inspired by the OPE in the following sense. The OPE constrains the singularities of $n$-point functions \emph{inside} of $\mathcal{W}$ in coincident limits to be controlled by other QFT operators. Since by our definition of $\mathcal{W}_0$, any point on $\partial \mathcal{W}$ is inside $\mathcal{W}_0$, if say a $1$-point function $\langle \Phi^{(i)}(x) \rangle$ diverges as $x\to \partial\mathcal{W}$, it is natural from the perspective of the OPE that this 1-point function can be captured as a higher point function in $\mathcal{W}_0$ where the divergence arises from approaching new operators (or their respective light cones) placed in $\mathcal{W}_0$.

\subsection{A construction of $\partial \mathcal{W}$ singularities with QFT operators}
\label{sec:32}

\begin{figure}[htbp]
\centering
\begin{minipage}{0.48\textwidth}
    \includegraphics[width=\linewidth]{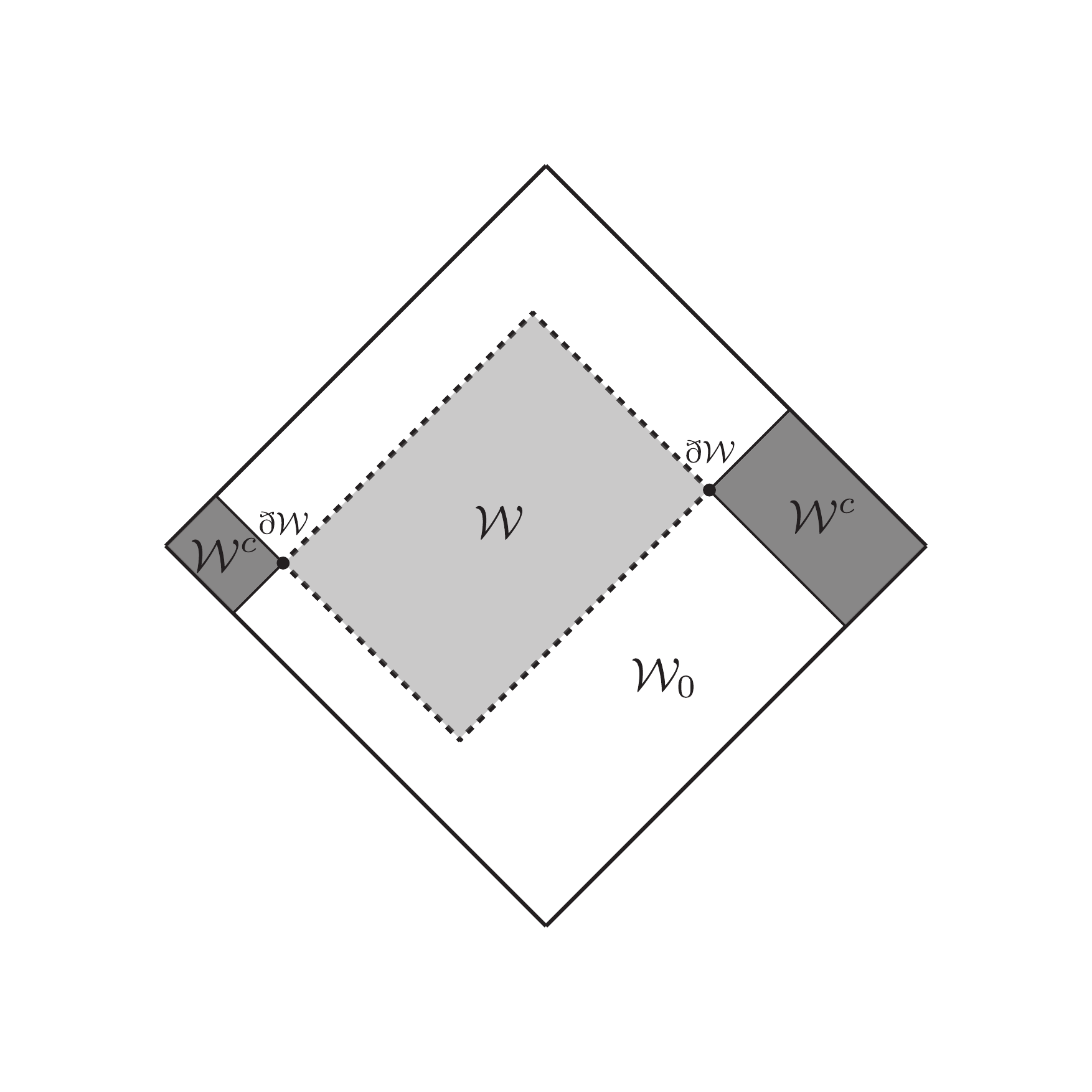} 
\end{minipage}
\hfill
\begin{minipage}{0.48\textwidth}
    \includegraphics[width=\linewidth]{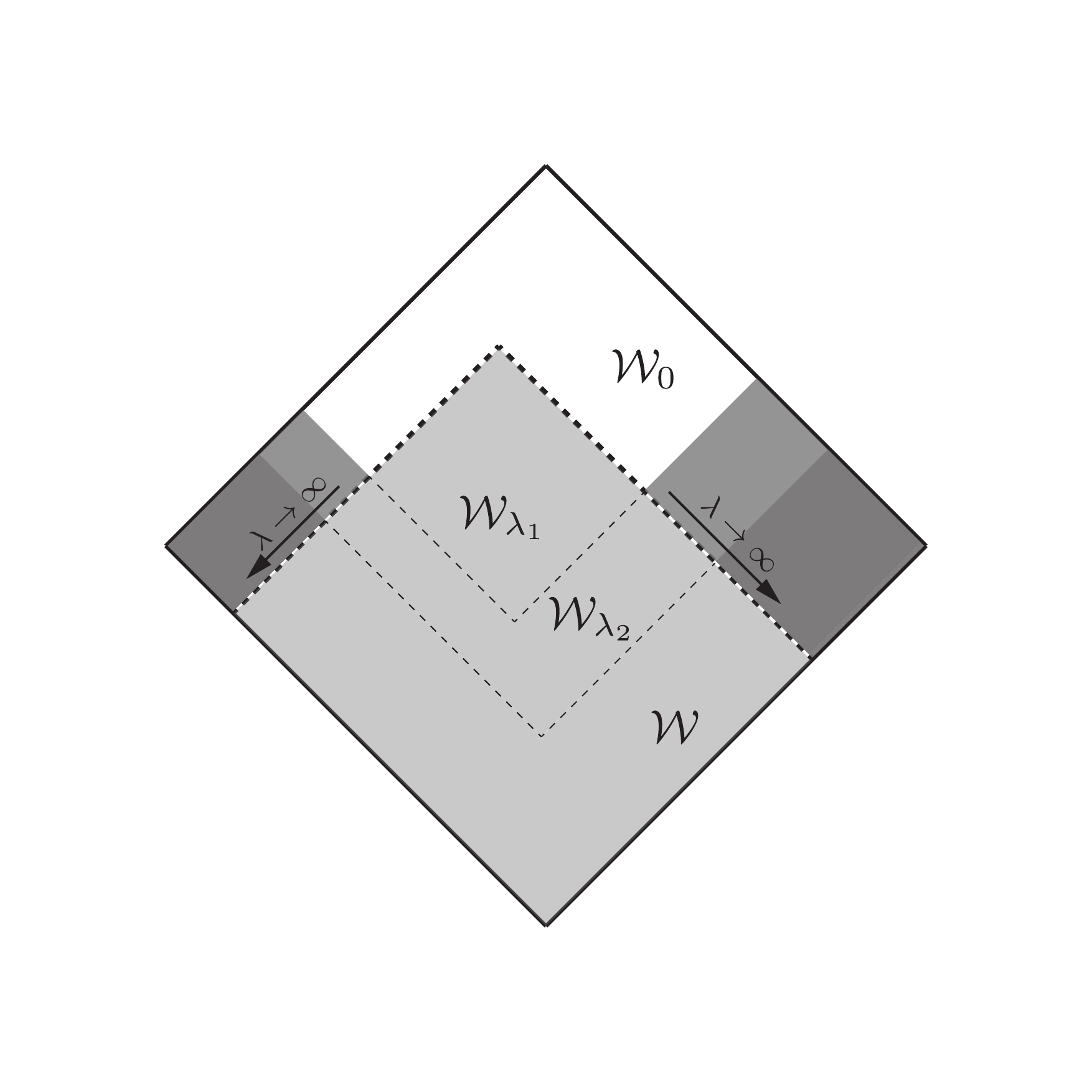} 
\end{minipage}
\caption{The left and right figures depict constructions \eqref{eq-precursor} and \eqref{eq-main1} respectively. In both figures, $\mathcal{W}$ is a wedge shown in grey inside a larger globally hyperbolic spacetime $\mathcal{W}_0$. On the left, $\mathcal{W}^c \neq \emptyset$ and is shown in darker grey. The edge of $\mathcal{W}$, denoted by $\eth\mathcal{W}$ is marked as two points. On the right, $\mathcal{W}^c = \emptyset$ and one can apply \eqref{eq-main1} by considering a family of growing wedges $\mathcal{W}_\lambda$, i.e., $\mathcal{W}_{\lambda_1} \subset \mathcal{W}_{\lambda_2}$ for $\lambda_1 < \lambda_2$ which limit to $\mathcal{W}$, and where for each $\lambda$ one simply applies construction \eqref{eq-precursor} depicted on the left.}
\label{fig3}
\end{figure}

Let us now restrict to $\mathcal{W}$ which is an open subset of a larger globally hyperbolic spacetime $\mathcal{W}_0$ in which it is maximally Cauchy evolved (see Figs.~\ref{fig3}), and further $\overline{\mathcal{W}} \subset \mathcal{W}_0$ (where $\overline{\mathcal{W}}$ denotes the closure of $\mathcal{W}$). We will refer to such $\mathcal{W}$ as a \emph{wedge}, and are interested in the allowed singularities on $\partial \mathcal{W}$, the boundary of $\mathcal{W}$ as a subset of $\mathcal{W}_0$. Even though by definition $\mathcal{W}$ is geometrically extendible to $\mathcal{W}_0$, it is not obvious that every consistent set of $n$-point functions in $\mathcal{W}$ can be extended to consistent $n$-point functions in $\mathcal{W}_0$, in particular, because even $1$-point functions of consistent $\mathcal{W}$ states are allowed to diverge on $\partial \mathcal{W} \subset \mathcal{W}_0$. However, any QFT must have at least one state $\omega_0$ in $\mathcal{W}_0$~\cite{Hollands:2009bke}. It is then plausible that any $\mathcal{W}$  state can be constructed from $\omega_0$ in some way. Though we will not be able to prove this latter statement, in the rest of this section we go the other way: we propose a construction of a broad class of states in $\mathcal{W}$ starting from a state in $\mathcal{W}_0$. Importantly, the resulting $\mathcal{W}$ states can have singularities on $\partial \mathcal{W}$.

From here on, we will denote QFT data in a wedge $\mathcal{W}$ associated to a state $\omega$ by the pair $(\mathcal{W},\omega)$. Obviously, any $(\mathcal{W}_0,\omega_0)$ automatically gives rise to a $(\mathcal{W},\omega)$, and one which has smooth $1$-point functions on $\partial \mathcal{W}$. The basic strategy to construct new $(\mathcal{W},\omega)$ from $(\mathcal{W}_0,\omega_0)$ is to let $n$-point functions in $\mathcal{W}$ be given by higher point functions in $\mathcal{W}_0$. For fixed 
\begin{align}
&\{j_1,\cdots,j_k\} \in I^k,\\
&\{y_1,\cdots,y_k\} \in (\mathcal{W}_0)^k,
\end{align}
let us define the following distribution $G: \mathcal{W}^n \to \mathbb{C}$:
\begin{align}
G(x_1,\cdots,x_n) = \langle \Phi^{(j_1)}(y_1) \cdots \Phi^{(j_k)}(y_k) \Phi^{(i_1)}(x_1) \cdots \Phi^{(i_n)}(x_n)\rangle^{\mathcal{W}_0}_{\omega_0}
\end{align}
The presence of the first $k$ operators may introduce singularities in $G$ which are not allowed by the MSC \emph{as a distribution on $\mathcal{W}^n$}. The full definition of MSC is cumbersome and though we will include it in appendix~\ref{sec:app1} for completeness, it is easy to explain a crucial aspect of it for our purposes. The main idea is that MSC constrains any singularity caused by an operator to propagate only within its domain of influence (the union of its causal future and past). Therefore, the only way to ensure that the singularities generated by operators $\Phi^{(j_1)}(y_1), \cdots, \Phi^{(j_k)}(y_k)$ do not violate the MSC for $G(x_1,\cdots,x_n)$ as a distribution on $\mathcal{W}$ is if $y_1, \cdots, y_k \in \mathcal{W}^c$, i.e. the causal complement of $\mathcal{W}$ within $\mathcal{W}_0$ (see appendix~\ref{sec:app1} for details). Note that we define the causal complement $\mathcal{W}^c$ as:
\begin{align}
\mathcal{W}^c = \{x \in \mathcal{W}_0 | J(x) \cap \mathcal{W}=\emptyset \}
\end{align}
where $J(x)$ denotes the domain of influence of $x$, i.e. the union of the causal past and future of $x$ (including $x$ itself). In particular, $\mathcal{W}$ will be closed in $\mathcal{W}_0$. See left Fig.~\ref{fig3}.

Guided by this, we define the following broad class of $(\mathcal{W}, \omega)$:
\begin{align}\label{eq-precursor}
\langle \Phi^{(i_1)}(x_1) \cdots \Phi^{(i_n)}(x_n)\rangle^{\mathcal{W}}_{\omega} = \langle \mathcal{D}(\mathcal{W}^c) \Phi^{(i_1)}(x_1) \cdots \Phi^{(i_n)}(x_n)\rangle^{\mathcal{W}_0}_{\omega_0}
\end{align}
where $\mathcal{D}(\mathcal{W}^c)$ is defined as follows:
\begin{align}\label{eq-b2x214c}
\mathcal{D}(\mathcal{W}^c) = \sum_{k=1}^\infty \sum_{\{j_1,\cdots, j_k\} \in I^k} \int_{\mathcal{W}^c} dy_1 \cdots \int_{\mathcal{W}^c} dy_k ~\tilde{f}^{(j_1 \cdots j_k)}(y_1,\cdots, y_k) \Phi^{(j_1)}(y_1) \cdots \Phi^{(j_k)}(y_k)
\end{align}
where $\tilde{f}^{(j_1,\cdots,j_k)}$ are a arbitrary functions on $(\mathcal{W}^c)^k$ such that the RHS of Eq. \eqref{eq-precursor} converges.\footnote{It is possible that even though each term in Eq.~\eqref{eq-b2x214c} gives a contribution to Eq.~\eqref{eq-precursor} consistent with the MSC, an infinite sum of such terms violate it. By restricting $\tilde{f}^{(j_1,\cdots,j_k)}$, we forbid such behavior in $\mathcal{D}(\mathcal{W}^c)$ by definition.} We furthermore allow $\tilde{f}^{(j_1,\cdots,j_k)}$ to be distributions with support contained within submanifolds of $(\mathcal{W}^c)^k$, so for example the operator $\mathcal{D}(\mathcal{W}^c)$ may have support only at a collection of points, lines, or more general embedded submanifolds of $(\mathcal{W}^c)^k$. We do, however, restrict to $\mathcal{D}(\mathcal{W}^c)$ such that the $n$-point functions on the LHS satisfy the positivity condition \eqref{eq-23cir15}, and that are unit normalized, i.e.,
\begin{align}
\langle \mathcal{D}(\mathcal{W}^c) \rangle^{\mathcal{W}_0}_{\omega_0} = 1.
\end{align}

These conditions ensure that the resulting $(\mathcal{W},\omega)$ satisfy the basic consistency conditions discussed earlier.\footnote{One may worry that even though the individual terms in the various sums in Eq. \eqref{eq-b2x214c} do respect the MSC, after the (possibly infinite) sum this is no longer the case. We will demand that this is in fact the case as another criteria in defining $\mathcal{D}(\mathcal{W}^c)$.} See Sec.~\ref{sec:33} for some examples. We will henceforth refer to operators $\mathcal{D}(\mathcal{W}^c)$ as \emph{defect operators}, due to analogy to other similarly defined operators in the literature (see e.g. \cite{Billo:2016cpy, Meineri:2016jpp, Balakrishnan:2017aa, Bousso:2014uxa, McAvity:1995zd, Cuomo:2021rkm}). In principle, we would like to expand the definition \eqref{eq-b2x214c} to include any operator-valued distribution with support contained in $\mathcal{W}^c$, even if they cannot be written as in Eq. \eqref{eq-b2x214c} in terms of other local operators.\footnote{The sense in which such an extended operator is localized can presumably be made concrete through the existence of an appropriate OPE when it is approached by local operators.} However, to our knowledge, the axiomatic approaches to QFT similar to~\cite{Hollands:2009bke} do not directly discuss such extended operator-valued distributions.\footnote{The existence of operator-valued distributions that cannot be written in terms of local operators is explicitly known in, for example, topological QFT. Therefore, we believe the lack of an explicit definition for such operators to be a shortcoming of the axiomatic frameworks like~\cite{Hollands:2009bke}.} Since we are working within that framework, we explicitly state only the very limited definition of $\mathcal{D}(\mathcal{W}^c)$ in Eq. \eqref{eq-b2x214c}, though, implicitly in what follows we allow arbitrary defect operators.

Let us make some clarifying remarks on the construction \eqref{eq-precursor}. In a QFT Hilbert space, new states can be obtained by acting with the smeared fields operator algebra $\mathcal{A}$ on some initial state. But importantly, acting with an operator-valued distribution does not define a new state. Therefore, the construction of \eqref{eq-precursor} should not be thought of as defining a new state in the GNS Hilbert space of $\omega_0$, but as defining a new set of $n$-point functions in $\mathcal{W}$ from which one can construct a \emph{new} Hilbert space according to the GNS construction. Such different Hilbert spaces constructed around the same background are often referred to as different \emph{sectors}. One can then explore different states within the same sector by acting with elements of the algebra $\mathcal{A}$ in $\mathcal{W}$.

The construction \eqref{eq-precursor} works when $\mathcal{W}^c \neq \emptyset$. This does not, however, hold for all $\mathcal{W} \subset \mathcal{W}_0$ (see right Fig.~\ref{fig3}). An example of this is $\mathcal{W}_P \subset \mathbb{R}^{1,d-1}$ which is of particular importance here. To deal with such cases, we generalize the construction as follows: Let $\mathcal{W}_\lambda$ be a one-parameter family of wedges for $\lambda \in [0,\infty)$, such that:
\begin{align}
&\mathcal{W}_{\lambda_1} \subseteq \mathcal{W}_{\lambda_2}, \hspace{2mm} \lambda_1 \leq \lambda_2\label{eq-2c41v}\\
&\mathcal{W} = \cup_\lambda \mathcal{W}_\lambda\label{eq-2c41v2}
\end{align}

We can generalize the construction of Eq. \eqref{eq-precursor} to
\begin{align}\label{eq-main1}
\langle \Phi^{(i_1)}(x_1) \cdots \Phi^{(i_n)}(x_n) \rangle^{\mathcal{W}}_\omega
= \lim_{\lambda \to \infty} \langle \mathcal{D}(\mathcal{W}_\lambda^c)~ \Phi^{(i_1)}(x_1) \cdots \Phi^{(i_n)}(x_n) \rangle^{\mathcal{W}_0}_{\omega_0},
\end{align}
for $x_1,\cdots,x_n \in \mathcal{W}$, where $\mathcal{D}(\mathcal{W}_\lambda^c)$ is a defect operator localized to $\mathcal{W}_\lambda^c$. Obviously, the construction requires that this limit exists for all $n$-point functions. Also, it should be clear that when all generators of $\partial\mathcal{W}$ have an endpoint on the \emph{edge} of $\mathcal{W}$, denoted by $\eth\mathcal{W}$, that Eq. \eqref{eq-main1} reduces to Eq. \eqref{eq-precursor}.\footnote{Recall that the edge of $\mathcal{W}$ is defined as the set of points $q$ in the closure of $\mathcal{W}$ such that every neighborhood of $q$ contains points $r_1$ and $r_2$ in the chronological future and past of $q$ respectively, and a timelike curve $\gamma$ connecting $r_1$ and $r_2$ such that $\gamma$ does not intersect the closure of $\mathcal{W}$~\cite{Wald:1984rg}.} (see Fig.~\ref{fig3}). In the next subsection, we will present several examples of this construction.

\subsection{Examples}
\label{sec:33}

Let us now see Eqs. \eqref{eq-precursor} and \eqref{eq-main1} in action with some basic examples. We will first focus on the case where there exists a $\mathcal{W}_0$ in which all the generators of $\partial \mathcal{W}$ have an endpoint on $\eth \mathcal{W}$. To create singularities on \emph{all} of $\partial \mathcal{W}$ starting from a smooth $(\mathcal{W}_0, \omega_0)$, the support of the defect operator must contain $\eth \mathcal{W}$. In particular we can consider a codimension-two defect $\mathcal{D}(\eth \mathcal{W})$ as a special case of $\mathcal{D} (\mathcal{W}^c)$, since $\eth \mathcal{W} \subset \mathcal{W}^c$.

For example, consider:
\begin{align}
\mathcal{D}(\mathcal{W}^c) = :\exp\left(\int_{\eth \mathcal{W}}d^2 y~ \phi^2\right):
\end{align}
where the colons indicate normal-ordering (see Appendix~\ref{sec:app3}). The exponential of an operator is to be understood as an expansion, like below:
\begin{align}\label{eq-124r31}
\mathcal{D}(\mathcal{W}^c)=1+ \sum_{m=1}^\infty \frac{1}{m!}\int_{\eth \mathcal{W}} dy^2_1 \cdots \int_{\eth \mathcal{W}} dy^2_m~ :\phi^2(y_1) \cdots \phi^2(y_m):
\end{align}
where the standard volume measure intrinsic to $\eth \mathcal{W}$ is used in the integral. The integrands in Eq. \eqref{eq-124r31} are operator-valued distributions and make sense inside correlation functions. The integral should be performed last.

To be more explicit, let us specialize to a free massless scalar theory, and further take $\mathcal{W} = \mathcal{W}_R$, $\mathcal{W}_0 = \mathbb{R}^{1,3}$, and let $\omega_0$ be the Minkowski vacuum state. Then, all generators of $\partial \mathcal{W}_R$ terminate on $\eth \mathcal{W}$, located at $U=V=0$. Now, a simple computation shows that construction \eqref{eq-precursor} with the defect \eqref{eq-124r31} results in the following $2$-point function in $\mathcal{W}_R$:
\begin{align}\label{eq-4552}
\langle \phi(x_1)\phi(x_2) \rangle^{\mathcal{W}_R}_\omega
=G_0(x_1,x_2) +\frac{1}{16 \pi ^3}\frac{\log(\frac{U_1 V_1}{U_2 V_2})}{U_1 V_1 - U_2 V_2}
\end{align}
where for simplicity we have placed both operators at the same transverse location $y^A$. Furthermore, $\omega$ is a Gaussian state of $\mathcal{W}_R$, i.e., all $n$-point functions with odd $n$ vanish and those with even $n$ are fixed by Eq. \eqref{eq-4552} in the following way:\footnote{We leave the demonstration of Gaussianity as an exercise.}
\begin{align}
\langle \phi(x_1) \cdots \phi(x_n) \rangle^{\mathcal{W}_R}_\omega = \sum_{\pi \in \Pi_{n}} \langle \phi(x_{\pi(1)}) \phi(x_{\pi(2)}) \rangle^{\mathcal{W}_R}_\omega \cdots \langle \phi(x_{\pi(n-1)}) \phi(x_{\pi(n)}) \rangle^{\mathcal{W}_R}_\omega
\end{align}
where $\Pi_{n}$ are the set of permutations of $\{1,\cdots, n \}$ such that $\pi(1) < \pi (3) < \cdots < \pi(n)$ and $\pi(1)<\pi(2), \cdots \pi(n-1) < \pi(n)$. Furthermore, all other $n$-point functions of fields are fixed by the $n$-point function of the fundamental fields and therefore, Eq. \eqref{eq-4552} defines the $\mathcal{W}_R$ state. Gaussian states are particularly abundant in physically relevant settings. For example, the HH and Unruh states of black holes are Gaussian states.

In Gaussian states, the MSC of \emph{all} $n$-point functions follows from the so-called Hadamard condition of the $2$-point function and the absence of any singularities at spacelike separated $x_1$ and $x_2$~\cite{Radzikowski:1996pa, Radzikowskilong, Hollands:2014eia, Kay:1988mu}. The Hadamard condition for any wedge in $\mathbb{R}^{1,d-1}$ is simply the condition that the $2$-point function is the sum of $G_0(x_1,x_2)$ and a smooth function of $x_1$ and $x_2$. The Hadamard and no spacelike singularity conditions are obviously satisfied by Eq. \eqref{eq-4552}.\footnote{Note that for simplicity we placed $x_1$ and $x_2$ on the same transverse position in Eq. \eqref{eq-4552}, so full smoothness and absence of spacelike singularities cannot be checked by looking at the equation, but it is an easy exercise which we leave to the reader.} The positivity condition for Gaussian states can also be reduced to the following requirement:
\begin{align}
\int_{\mathcal{W}_R}\int_{\mathcal{W}_R} d^4x_1 d^4x_2~f(x_1) \bar{f}(x_2) \langle \phi(x_1)\phi(x_2) \rangle^{\mathcal{W}_R}_\omega \geq 0
\end{align}
where $f(x)$ is a smearing function and $\bar{f}(x)$ is its complex conjugate~\cite{Hollands:2014eia}. This condition is also obviously satisfied by Eq. \eqref{eq-4552} since it is the sum of a positive smooth function and $G_0(x_1,x_2)$, the (positive) Minkowski vacuum $2$-point function. Therefore, the state \eqref{eq-4552} satisfies the QFT constraints emphasized in Sec.~\ref{sec:31}. The argument for compatibility with QFT axioms goes through very similarly for the rest of the example states in this section which are Gaussian, so we will not repeat them.

From Eq. \eqref{eq-4552}, we can derive the following $1$-point functions:
\begin{align}
&\langle \phi^2(x) \rangle^{\mathcal{W}_R}_\omega = \frac{1}{16 \pi^3 UV}\\
&\langle T_{VV}(x) \rangle^{\mathcal{W}_R}_\omega =  \frac{1}{48 \pi^3 UV^3}
\end{align}
which are singular at $\partial \mathcal{W}$.

Another family of states in $\mathcal{W}_R$ with $\partial \mathcal{W}_R$ singularities are the ``wrong temperature'' states which, as discussed in Sec.~\ref{sec:2}, can be constructed by inserting a conical singularity at $\eth \mathcal{W}_R$ in the Euclidean section. This can be viewed as the insertion of a so-called twist defect at $\eth \mathcal{W}_R$ (see~\cite{Calabrese:2004eu} for a detailed discussion). For a free theory, this results in Gaussian states with explicitly known two-point functions of which we have already included an example in Eq.~\eqref{eq-13e1f4}. As such, these states also serve as an example of Eq. \eqref{eq-precursor}.

Now, let us provide some examples of the \eqref{eq-main1} construction with $\mathcal{W}_0 = \mathbb{R}^{1,3}$ and $\omega_0$ the Minkowski vacuum as above, but where $\mathcal{W}^c = \emptyset$. The past Rindler wedge $\mathcal{W}_P$ as a subset of Minkowski space is a standard example of this. Let $L$ be an arbitrary length-scale. We choose the following:
\begin{align}\label{eq-21i4crni214r}
\mathcal{W}_\lambda = \{ -\lambda L <U< 0, -\lambda L<V< 0 \}
\end{align}
Then we have that $\eth \mathcal{W}_\lambda = \mu_+(\lambda) \cup \mu_-(\lambda)$ where:
\begin{align}\label{eq-347r3}
&\mu_+(\lambda) = \{(U=-\lambda L,V=0, y^A \in \mathbb{R}^2)\}\\
&\mu_-(\lambda) = \{(U=0,V=-\lambda L, y^A \in \mathbb{R}^2)\}
\end{align}
As a first example, consider the defect
\begin{align}\label{eq-2i4f1n09132yc}
\mathcal{D}(\mathcal{W}^c)= :\exp{\left(\int_{\mu_+(\lambda)} d^2 y~\partial_V \phi\right)}:
\end{align}
Note that this defect acts as the identity operator on the $\mu_-(\lambda)$. The resulting $\mathcal{W}_P$ state is not Gaussian. Its $n$-point functions can be derived by performing the following operator replacement inside Minkowski vacuum correlation functions:
\begin{align}\label{eq-21c3415c}
\phi(x) \to \phi(x) - \frac{1}{4 \pi V} \mathbb{1}
\end{align}
The replacement \eqref{eq-21c3415c} should make explicit the positivity of the state and its compatibility with MSC.\footnote{This state is very simply related to the (Gaussian) Minkowski vacuum state, so we leave the checking of QFT axioms as an exercise. But since this state is not Gaussian, the arguments for positivity and compatibility with MSC from earlier in the section do not directly apply.} In particular, the $2$-point function is given by:
\begin{align}
\langle \phi(x_1)\phi(x_2) \rangle^{\mathcal{W}_P}_\omega
=G_0(x_1,x_2)+\frac{1}{16 \pi^2 V_1 V_2}
\end{align}

This state has also singularities in $1$-point functions. For example:
\begin{align}
&\langle \phi^2(x) \rangle^{\mathcal{W}_P}_\omega = \frac{1}{16\pi^2 V^2}\\
&\langle T_{VV}(x) \rangle^{\mathcal{W}_P}_\omega =\frac{1}{16 \pi^2 V^4}
\end{align}
Note that this state does not have singularities on the $U=0$ portion of $\partial \mathcal{W}_P$. This follows from the defect \eqref{eq-2i4f1n09132yc} being spacelike separated from this portion of the horizon.

From this, it is trivial to construct an example of a defect which causes singularities on all of  $\partial \mathcal{W}_P$. For example:
\begin{align}
\mathcal{D}(\mathcal{W}_\lambda^c) = \frac{1}{2} \left[: \exp \left(\int_{\mu_+(\lambda)}d^2 y~\partial_V \phi \right): + : \exp \left(\int_{\mu_-(\lambda)}d^2 y~\partial_U \phi \right): \right]
\end{align}

In a conformal compactification of Minkowski space, all generators of the $V=0$ ($U=0$) part of $\partial \mathcal{W}_P$ ends on the same point $p_+$ ($p_-$) on the past null infinity. In fact, $\mathcal{W}_P$ is obtained by merely removing the future light cones of $p_+$ and $p_-$ from Minkowski spacetime! See subsection~\ref{sec:41} for more details on this aspect of the $\mathcal{W}_P$ geometry.

The following three defect highlights this feature of $\mathcal{W}_P$. First, consider
\begin{align}\label{eq-b7n87b1v}
\mathcal{D}(\mathcal{W}_\lambda^c) =~ :\exp{\left(\lambda^2 L^4 T_{VV}(p^+_\lambda)\right)}:~:\exp{\left(\lambda^2 L^4 T_{UU}(p^-_\lambda)\right)}:
\end{align}
where
\begin{align}
&p^\lambda_+ = (U=-\lambda L, V=0, y_+^A),\\
&p^\lambda_- = (U=0, V=-\lambda L, y_-^A),
\end{align}
where $y_+^A$ and $y_-^A$ can be chosen arbitrarily. The missing multiplicative normalization factor in Eq. \eqref{eq-b7n87b1v} is obtained by demanding $\langle \mathcal{D}(\mathcal{W}_\lambda^c) \rangle^{\mathcal{W}_0}_{\omega_0}=1$.
The points $p^\lambda_\pm$ limit to $p_\pm$ on the conformal boundary as $\lambda \to \infty$. Following Eq.~\eqref{eq-main1}, the defect \eqref{eq-b7n87b1v} gives rise to a Gaussian state in $\mathcal{W}_P$ with the following $2$-point function:
\begin{align}
\langle \phi(x_1)\phi(x_2) \rangle^{\mathcal{W}_P}_\omega
=G_0(x_1,x_2) + \frac{L^2}{16 \pi^4 V_1^2 V_2^2}+ \frac{L^2}{16 \pi^4 U_1^2 U_2^2}
\end{align}
with in particular the following one-point function singularities:
\begin{align}
\langle \phi^2(x) \rangle^{\mathcal{W}_P}_\omega &= \frac{L^2}{16 \pi^4 V^4} +\frac{L^2}{16 \pi^4 U^4}, \\
\langle T_{VV}(x) \rangle^{\mathcal{W}_P}_\omega &=\frac{L^2}{16 \pi^4 V^6},\\
\langle T_{UU}(x) \rangle^{\mathcal{W}_P}_\omega &=\frac{L^2}{16 \pi^4 U^6}.
\end{align}
Therefore, we see that even though the defect Eq. \eqref{eq-nbbwv9322} only has support over $p^\lambda_\pm$ which asymptote to the two points $p_\pm$ on the conformal boundary as $\lambda \to \infty$, it gives rise to a state which has singularities on all of $\partial \mathcal{W}_P$.
The defect \eqref{eq-b7n87b1v} can be factorized into a product of a pair of defects localized to $p^\lambda_+$ and $p^\lambda_-$. This need not be necessarily the case for all defects. For example, consider
\begin{align}\label{eq-nbbwv9322}
D(\mathcal{W}_\lambda^c) =~:\exp{\left(\lambda^2 L^2 \phi^2(p^+_\lambda)\right)} ~\exp{\left(\lambda^2 L^2 \phi^2(p^-_\lambda)\right)} :
\end{align}
Despite the factorized form between the colons, the normal ordering introduces terms in this defect which couple points in $p^+_\lambda$ and $p^-_\lambda$. This becomes clear after expanding the first few terms in the defect \eqref{eq-nbbwv9322} explicitly:
\begin{align}
D(\mathcal{W}_\lambda^c) &= 1 + \lambda^2 L^2 :\phi^2(p^+_\lambda): + \lambda^2 L^2 :\phi^2(p^-_\lambda):- 4\lambda^4 L^4 \langle \phi(p^+_\lambda) \phi(p^-_\lambda) \rangle^{\mathcal{W}_0}_{\omega_0} ~\phi(p^+_\lambda) \phi(p^-_\lambda)\nonumber\\
&-\lambda^4 L^4 \langle :\phi^2(p^+_\lambda): :\phi^2(p^-_\lambda): \rangle^{\mathcal{W}_0}_{\omega_0}~\mathbb{1} + \cdots.
\end{align}
This defect results in a Gaussian state of $\mathcal{W}_P$ with $2$-point function
\begin{align}\label{eq-2c1v55v6b}
\langle \phi(x_1)\phi(x_2) \rangle^{\mathcal{W}_P}_\omega
=G_0(x_1,x_2) + \frac{1}{16 \pi^4 V_1 V_2}+ \frac{1}{16 \pi^4 U_1 U_2},
\end{align}
and the following $1$-point functions:
\begin{align}
\langle \phi^2(x) \rangle^{\mathcal{W}_P}_\omega &= \frac{1}{16 \pi^4 V^2} +\frac{1}{16 \pi^4 U^2}, \\
\langle T_{VV}(x) \rangle^{\mathcal{W}_P}_\omega &=\frac{1}{16 \pi^4 V^4}.
\end{align}

We can also consider defects which couple the points $p^+_\lambda$ and $p^-_\lambda$ even before normal-ordering. For example, consider
\begin{align}\label{eq-347h191}
D(\mathcal{W}_\lambda^c) =~ :\exp{\left(\lambda^2 L^2\phi(p^+_\lambda)\phi(p^-_\lambda) \right)}:
\end{align}
This results in a Gaussian state with the following $2$-point function:
\begin{align}\label{eq-24v23vb77b}
\langle \phi(x_1)\phi(x_2) \rangle^{\mathcal{W}_P}_\omega
=G_0(x_1, x_2) +\frac{1}{16 \pi^4 V_1 U_2} + \frac{1}{16 \pi^4 V_2 U_1},
\end{align}
leading to the following $1$-point function singularity:
\begin{align}\label{eq-2v45}
\langle \phi^2(x) \rangle^{\mathcal{W}_P}_\omega = \frac{1}{8\pi^4 UV}.
\end{align}
This examples demonstrates a significant point. The state \eqref{eq-24v23vb77b} is $k^\mu$-symmetric and the singularity of Eq. \eqref{eq-2v45} is fully commensurate with our philosophy outlined in the introduction. Namely, it is predictable as a leading singularity subject to symmetry and dimensional analysis. It therefore shows that the absence of such singularities in the numerical analysis of black hole interiors (see e.g.~\cite{Zilberman:2019buh} where $\phi^2$ was found to be regular at the Cauchy horizon) cannot simply be because this is not allowed on completely general grounds. In Sec.~\ref{sec:4}, we invoke an additional criteria (namely, the robustness of a singularity) which rules out the singularity of \eqref{eq-2v45}.

The defects discussed so far only have support contained within the edge of $\mathcal{W}^c_\lambda$. This need not be the case. For example, consider:
\begin{align}
&D(\mathcal{W}_\lambda^c) \sim~:\exp{\left[\lambda^2 \left(\int_0^L dV' V' \partial_V \phi(U=-\lambda L, V',y_+^A)\right)^2 \right]}:\nonumber\\
&\hspace{5cm} \times:\exp{\left[\lambda^2 \left( \int_0^L dU' U'\partial_U \phi(U', -\lambda L,y_-^A)\right)^2 \right]} :
\end{align}
It is a simple exercise to show that this defect results in a Gaussian $\mathcal{W}_P$ state with the following 2-point function:
\begin{align}
\langle \phi(x_1)\phi(x_2) \rangle^{\mathcal{W}_P}_\omega
&=G_0(x_1, x_2) \nonumber\\
&+\frac{1}{8 \pi^4 L^2} \left(\log \left(\frac{L+V_1}{V_1}\right) + \frac{L}{L+V_1}\right)\left(\log \left(\frac{L+V_2}{V_2}\right) + \frac{L}{L+V_2}\right) \nonumber\\
&+ \frac{1}{8 \pi^4 L^2} \left(\log \left(\frac{L+U_1}{U_1}\right) + \frac{L}{L+U_1} \right)\left(\log \left(\frac{L+U_2}{U_2}\right) + \frac{L}{L+U_2} \right),
\end{align}
with the following $1$-point function singularities as operators are taken to $V=0$ at fixed $U<0$:
\begin{align}
\langle \phi^2(x) \rangle^{\mathcal{W}_P}_\omega &\stackrel{V\to0^-}{=} \frac{1}{8 \pi^4 L^2} \left(\log \left(\frac{L}{V} \right)\right)^2+\cdots, \\
\langle T_{VV}(x) \rangle^{\mathcal{W}_P}_\omega &\stackrel{V\to0^-}{=} \frac{1}{8 \pi^4 L^2 V^2}+\cdots.
\end{align}
where ellipsis denotes subleading corrections in the $V \to 0^-$ limit.

Here, we merely discussed some very simple example of defects in a free scalar theory to demonstrate the constructions of subsection~\ref{sec:32} in action. In this next section, we adopt $\mathcal{W}_P$ states constructed via construction Eq.~\eqref{eq-main1} with singularities on $\partial \mathcal{W}_P$ as models of black hole interior states with Cauchy horizon singularities.

\section{The Singularity Structure of $\mathcal{W}_P$ States}
\label{sec:4}

The discussion of Sec.~\ref{sec:3} was, on the surface, detached from black hole interiors and their Cauchy horizons. However, any black hole Cauchy horizon in the ``near Cauchy horizon limit'' approaches $\mathcal{W}_P$, and its state in that strict limit, results in a $\mathcal{W}_P$ state. Since the construction \eqref{eq-main1} is quite general, we find it plausible that all such $\mathcal{W}_P$ states can be constructed as special cases of Eq.~\eqref{eq-main1}. As we have already seen, the ``wrong temperature'' states with singular outer horizons (discussed in Sec.~\ref{sec:2}) already are special cases of our construction applied to $\mathcal{W}_R$. Here, we assume the same effectiveness for the construction \eqref{eq-main1} when applied to $\mathcal{W}_P$. Concretely, in Sec.~\ref{sec:42} we define \emph{defect constructible} states as a class of $\mathcal{W}_P$ states constructible via Eq.~\eqref{eq-main1}. Remarkably, we find that these states satisfy a mild singularity ansatz reminiscent of the black Cauchy horizon. We find this supporting evidence that indeed restricting to defect constructible states as models of black hole Cauchy horizon singularities is legitimate.

A plausible explanation for the effectiveness (in reproducing a mild singularity ansatz) of ``defect constructible'' states would be if \emph{all} $\mathcal{W}_P$ states are in fact ``defect constructible''. While we do not claim this, in the next subsection, we show that a very general class of conformal field theory (CFT) states in $\mathcal{W}_P$ indeed are \emph{all} special cases of construction~\eqref{eq-main1} and are therefore ``defect constructible''. This further justifies our restricting to ``defect constructible'' states.

\subsection{A class of CFT states in $\mathcal{W}_P$ descend from construction \eqref{eq-main1}}\label{sec:41}

\begin{figure}[htbp]
\centering
\includegraphics[width=0.6\textwidth]{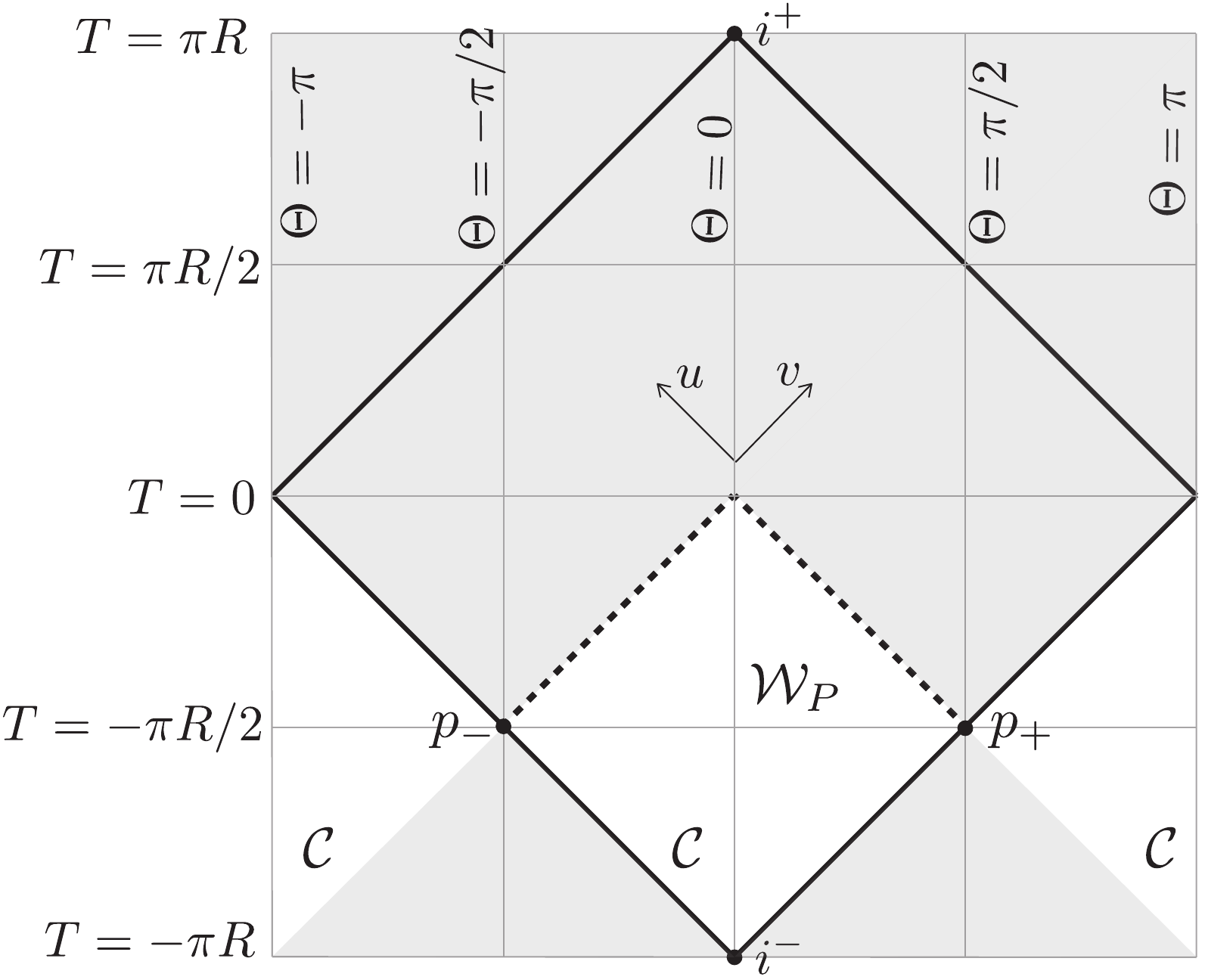} 
\caption{Embedding of Minkowski spacetime (large diamond given by $I^-(i^+) \cap I^+(i^-)$) into the cylinder $\mathcal{C}_0$ (large square) according to Eqs.~\eqref{eq-Weyl}, \eqref{eq-Weyl2}, and \eqref{eq-Weyl3}, with the $\Omega_{d-2}$ directions suppressed. The leftmost and rightmost vertical lines of the square ($\Theta=\pi$ and $\Theta=-\pi$ lines) are identified. The points $p_\pm$ on the past null infinity of Minkowski space are shown at $T=-\pi R/2$ and $\Theta = \pm \pi /2$ of $\mathcal{C}_0$. The grey region denotes $J(p_+) \cup J(p_-)$ which if removed from $\mathcal{C}_0$ results in $\mathcal{C}$. The region $\mathcal{W}_P$ (small diamond) is an even smaller region, though any analytic CFT state of $\mathcal{W}_P$ which is a restriction of a $\mathcal{C}$ state comes from inserting a bilocal defect $\tilde{\mathcal{D}}(p_+,p_-)$ into a state of $\mathcal{C}_0$ according to Eq. \eqref{eq-precursor}.}
\label{fig4}
\end{figure}

We will first discuss a conformal compactification of $\mathcal{W}_P$, and show that a special class of CFT states provably descend from of Eq.~\eqref{eq-main1}. As discussed before, $\mathcal{W}$ can be obtained by erasing two asymptotic points, $p_+$ and $p_-$, ``worth of information'' from $\mathbb{R}^{1,d-1}$. To quantify this more precisely, it will be useful to consider the following standard conformal compactification of $\mathbb{R}^{1,d-1}$ via the following Weyl transformation:
\begin{align}\label{eq-Weyl}
&ds^2 = \Omega^2 \left(-du dv + \frac{1}{4}(v-u)^2 d\Omega_{d-2}^2\right),\\
&\Omega^2 = \frac{4}{(1+v^2/R^2)(1+u^2/R^2)},
\end{align}
where $d\Omega_{d-2}^2$ is the metric of a unit $(d-2)$-sphere and $R$ is an arbitrary length scale. Note the distinction between this and the $(U,V,y^A)$ Minkowski coordinates used in Eq. \eqref{eq-2c124523}. After the changes of variables
\begin{align}
&T = R\arctan \left(\frac{v}{R}\right) + R\arctan \left(\frac{u}{R}\right),\label{eq-Weyl2}\\
&\Theta = \arctan \left(\frac{v}{R}\right) - \arctan \left(\frac{u}{R}\right),\label{eq-Weyl3}
\end{align}
the metric becomes:
\begin{align}
ds^2 = -dT^2 + R^2 d\Theta^2 + R^2 \sin^{2}\Theta ~d\Omega_{d-2}^2.
\end{align}
The conclusion is that the Weyl transformation \eqref{eq-Weyl} of $\mathbb{R}^{1,d-1}$ embeds it into a submanifold of a cylinder $\mathcal{C}_0=\mathbb{R} \times S^{d-1}$ with radius $R$. This submanifold is $I^+(i^-) \cap I^-(i^+)$ where $I^\pm$ denotes chronological future/past and
\begin{align}
i^\pm = (T=\pm \pi R, \Theta = 0)
\end{align}
are the future and past timelike infinities of $\mathbb{R}^{1,d-1}$(see Fig.~\ref{fig4}). Comparatively, $\mathcal{W}_P$ corresponds to an even smaller portion of $\mathcal{C}_0$. In fact, we can embed $\mathcal{W}_P$ into $\mathcal{C}_0$ minus ``two points worth of data'', by which we mean there exists an embedding map $F$ such that:
\begin{align}\label{eq-weylF}
F:\mathcal{W}_P \to \mathcal{C}_0 - J(p_+ \cup p_-).
\end{align}
Here, $p_{\pm} =  (T=-\pi R/2, \Theta = \pm \pi/2)$ and $J (.)$ denotes the causal future and past (including the points in the argument). Therefore, $F(\mathcal{W}_P)$ arises from Cauchy evolving (forwards and backwards) the $T=-\pi R/2$ sphere of $\mathcal{C}_0$ minus only two anti-podal points. For brevity, we henceforth use $\mathcal{C} = \mathcal{C}_0- J(p_+ \cup p_-)$. See Fig.~\ref{fig4}.

In a CFT, $n$-point functions on $\mathcal{W}_P$ are related to $n$-point functions in $F(\mathcal{W}_P)$ in the following way:
\begin{align}\label{eq-243c124}
\langle \tilde{\Phi}^{(i_1)}(x_1) \cdots\tilde{\Phi}^{(i_n)}(x_n)\rangle_{\tilde{\omega}}^{F(\mathcal{W}_P)} =  \langle \Phi^{(i_1)}(x_1) \cdots \Phi^{(i_n)}(x_n) \rangle_{\omega}^{\mathcal{W}_P}
\end{align}
where the local operators $\tilde{\Phi}^{(i)}$ in $\mathcal{C}$ are related by known Weyl transformations to $\mathcal{W}_P$ local operators $F(\mathcal{W}_P)$~\cite{DiFrancesco:1997nk, Simmons-Duffin:2016gjk, Farnsworth:2017tbz}, and $\tilde{\omega}$ is an appropriately ``Weyl-transformed'' state. For example, given a \emph{primary} CFT scalar operator $\mathcal{O}$ of dimension $\Delta$, we have:
\begin{align}
\tilde{\mathcal{O}}(x) = \Omega(x)^{-\Delta} \mathcal{O} (x) + \text{Weyl anomaly}
\end{align}
where $\Omega$ is the Weyl factor in Eq. \eqref{eq-Weyl}. The ``Weyl anomaly'' term can in principle contain other operators, though for a generic CFT we only expect such terms for certain primaries, such as $T_{\mu\nu}$ in even spacetime dimensions where the Weyl anomaly consists of a local geometric term multiplying the identity operator~\cite{Farnsworth:2017tbz}.

We will now restrict to states on $\mathcal{W}_P$ whose $n$-point functions, after the Weyl transformation \eqref{eq-243c124} can be extended to the larger spacetime $\mathcal{C}$. Since $\partial \mathcal{W}_P \subset \partial \mathcal{C}$, this class of states ought to exhibit a non-trivial singularity structure on $\partial \mathcal{W}_P$, though they do not constitute the most general analytic states on $\mathcal{W}_P$. In this class of states, we can (using the Weyl transformation) convert $\mathcal{W}_P$ $n$-point functions to $n$-point functions:
\begin{align}
\langle \tilde{\Phi}^{(i_1)}(x_1) \cdots\tilde{\Phi}^{(i_n)}(x_n)\rangle^{\mathcal{C}}_{\tilde{\omega}}.
\end{align}
Let us now place these $\mathcal{C}$ $n$-point functions on the time slice $T = -\pi R/2$, and continue them (for each $x_i$) to 
\begin{align}
T_i = -\frac{\pi}{2}R + i \tau_i,
\end{align}
Since the missing points $p_{\pm}$ live at $T = -\pi R/2$, this defines $n$-point functions of a Euclidean CFT on $(\mathbb{R}_E \times \mathbb{S}^{d-1}) - p_1-p_2$ (the subscript $E$ emphasizes Euclidean time), where these $n$-point functions can have singularities at $p_\pm$. At this stage, we take advantage of the \emph{state-operator correspondence} in the CFT to constrain the form of these singularities~\cite{DiFrancesco:1997nk, Simmons-Duffin:2016gjk}. In particular, in the Euclidean spacetime, consider a pair of very small $(d-1)$-dimensional spheres cut around points $p_+$ and $p_-$. The state on the union of these two spheres can be written in terms of an entangled sum over energy eigenstates of the CFT on each $(d-1)$-dimensional sphere.  The state-operator correspondence then instructs us that these energy eigenstates must arise from placing a possibly infinite but convergent sum of local primary operators and their descendants at $p_+$ and $p_-$. This constitutes a defect operator, denoted by $\tilde{\mathcal{D}} (p_+, p_-)$, with support on both $p_+$ and $p_-$:
\begin{align}\label{eq-12c5v}
\tilde{\mathcal{D}} (p_+, p_-) = \sum_{i_1 i_2} c_{i_1 i_2} \tilde{\Phi}^{(i_1)}(p_+) \tilde{\Phi}^{(i_2)}(p_-)
\end{align}
where $c_{i_1 i_2}$ are complex numbers. We can now write:
\begin{align}\label{eq-31c12}
\langle \tilde{\Phi}^{(i_1)}(x_1) \cdots\tilde{\Phi}^{(i_n)}(x_n)\rangle^{\mathcal{C}}_{\tilde{\omega}} = \langle \tilde{\mathcal{D}} (p_+, p_-) \tilde{\Phi}^{(i_1)}(x_1) \cdots\tilde{\Phi}^{(i_n)}(x_n)\rangle^{\mathcal{C}_0}_{\tilde{\omega}_0}.
\end{align}
where $\tilde{\mathcal{D}}(p_+, p_-)$ is normalized so that $\langle 1 \rangle^{\mathcal{C}}_\omega  = 1$. Eq. \eqref{eq-31c12} is an example of construction~\eqref{eq-precursor} (which in turn is a special case of Eq.~\eqref{eq-main1}) for the specific choices $\mathcal{W}=\mathcal{C}$, and $\mathcal{W}_0=\mathcal{C}_0$. 

Having derived Eq. \eqref{eq-31c12}, we can derive the corresponding $(\mathcal{W}_P, \omega)$ data via the Weyl transformation of $n$-point functions:

\begin{align}\label{eq-23rrc1}
\langle \Phi^{(i_1)}(x_1) \cdots \Phi^{(i_n)}(x_n)\rangle^{\mathcal{W}_P}_\omega = \langle \tilde{\mathcal{D}} (p_+, p_-) \tilde{\Phi}^{(i_1)}(x_1) \cdots\tilde{\Phi}^{(i_n)}(x_n)\rangle^{\mathcal{C}}_{\tilde{\omega}}.
\end{align}
The $\mathcal{W}_P$ state \eqref{eq-23rrc1} may have singularities on $\partial \mathcal{W}_P$ depending on the choice of $\tilde{\mathcal{D}} (p_+, p_-)$. In fact, it is a straightforward exercise to show that all of the defects in Sec.~\ref{sec:33} which are supported on points are examples of $\tilde{\mathcal{D}} (p_+, p_-)$. However, other defects whose support is larger than the two points the the $\lambda \to \infty$ limit approach a larger subset of past null infinity (larger than just $p_+ \cup p_-$). This support however, is still constrained to have a causal future which does not intersect $\mathcal{W}_P$. It can be shown that the largest subset of the past null infinity with this property is the union of the two future-inextendible null generators of past null infinity that emanate from $p_+$ and $p_-$ (denoted by $\mathcal{W}^+_\infty$ and $\mathcal{W}^-_\infty$ respectively in the next subsection). Therefore, a more general defect will approach $\mathcal{W}^+_\infty \cup \mathcal{W}^-_\infty$.

\subsection{The singularity structure of defect constructible states in $\mathcal{W}_P$}\label{sec:42}

We now restrict to the following class of states:

\begin{definition}\label{def1}
For any QFT in $\mathcal{W}_P$, we call a state $\omega$ ``defect constructible'' if there exists a $\mathbb{R}^{1,d-1}$ state $\mathcal{\omega}_0$ and the family of wedges,
\begin{align}\label{eq-b765}
\mathcal{W}_\lambda = \{ -\lambda L <U< 0, -\lambda L<V< 0 \},
\end{align}
where $L$ is an arbitrarily length scale, and a choice of defect operator $\mathcal{D}(\mathcal{W}_\lambda^c)$ such that

\begin{align}\label{eq-assumption1}
 \langle \Phi^{(i_1)}(x_1) \cdots \Phi^{(i_n)}(x_n) \rangle^{\mathcal{W}_P}_\omega
= \lim_{\lambda \to \infty}\langle \mathcal{D}(\mathcal{W}_\lambda^c) \Phi^{(i_1)}(x_1) \cdots \Phi^{(i_n)}(x_n) \rangle^{\mathbb{R}^{1,d-1}}_{\omega_0}
\end{align}
\end{definition}
Since $(\mathcal{W},\omega)$ is allowed to have singularities on $\partial \mathcal{W}$, but $(\mathcal{W}_0,\omega_0)$ does not, Eq. \eqref{eq-assumption1} in particular implies that any $\partial \mathcal{W}$ singularity is ``caused'' by the defect operators $\mathcal{D}(\mathcal{W}_\lambda^c)$.

We reiterate that the given the discussion of the previous section, it is reasonable to wonder whether \emph{all} $\mathcal{W}_P$ states are constructible via Eq. \eqref{eq-assumption1}. This is an attractive possibility, and if true, would make the above restriction unnecessary for what follows. Since we cannot prove this here, we simply employ this restriction. Further investigation of this will be left to future work. Next, we will constrain the singularity structure of $\mathcal{W}_P$ for defect constructible states. Along the way, we make assumptions which we state clearly.

Let $\omega$ be a defect constructible state in $\mathcal{W}_P$. Suppose we have a field $\Phi^{(i)}(x)$ with $x \in \mathcal{W}_P$ such that
\begin{align}\label{eq-234d2c2}
\lim_{V\to0^-} |\langle \Phi^{(i)}(U,V,y^A) \rangle^{\mathcal{W}_P}_\omega| = \infty,
\end{align}
where the vertical lines denote absolute value. We assume that any such singularity admits an asymptotic expansion
\begin{align}\label{eq-89h4925g}
\langle \Phi^{(i)}(U,V,y^A) \rangle^{\mathcal{W}_P}_\omega \stackrel{V\to0^-}{=} \sum_{n=0}^\infty (P^{(i)}_{\omega})^{(n)}(U,V,y^A)
\end{align}
where at least $(P^{(i)}_{\omega})^{(n)}$ are smooth function in a neighborhood of $V<0$ in $\mathcal{W}_P$. By an asymptotic expansion, we mean that for any fixed $U$ and $y^A$, and for any $N$, we have:
\begin{align}
 \langle\Phi^{(i)}(U,V,y^A) \rangle - \sum_{n=0}^N (P^{(i)}_{\omega})^{(n)}(U,V,y^A) \stackrel{V\to0^-}{=} o\left((P^{(i)}_{\omega})^{(N)}(U,V,y^A)\right).
\end{align}
We are primarily interested in the leading singularity behavior which is governed by $(P^{(i)}_\omega)^{(0)}$. We define the following notion:

\begin{definition}\label{def2}
An operator $\Phi^{(i)}$ satisfying Eq. \eqref{eq-234d2c2} has a \emph{robust singularity} if and only if:
\begin{align}\label{eq-h98g89y7}
\langle \Phi^{(i)}(U,V,y^A) \rangle^{\mathcal{W}_P}_{\omega'} \stackrel{V\to 0^-}= (P^{(i)}_\omega)^{(0)} (U, V, y^A) + \sum_{n=1}^\infty (P^{(i)}_{\omega'})^{(n)}(U,V,y^A)
\end{align}
where $\omega'$ is any state in the GNS Hilbert space of $\omega$.
\end{definition}
Note that the robustness of a singularity is a feature of the Hilbert space, and not the specific state under consideration.

\begin{figure}[htbp]
\centering
\begin{minipage}{0.48\textwidth}
    \includegraphics[width=\linewidth]{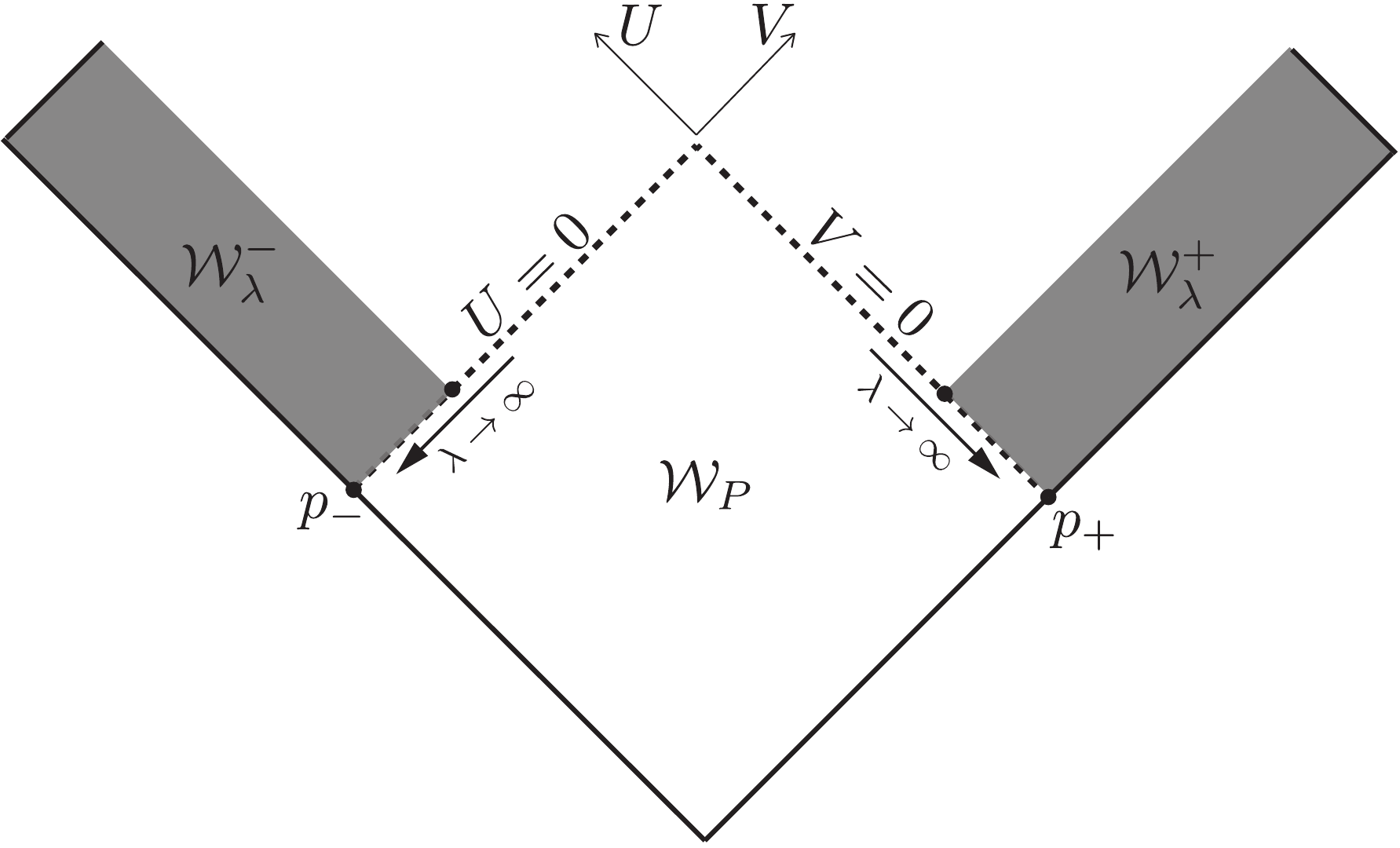} 
\end{minipage}
\hfill
\begin{minipage}{0.48\textwidth}
    \includegraphics[width=\linewidth]{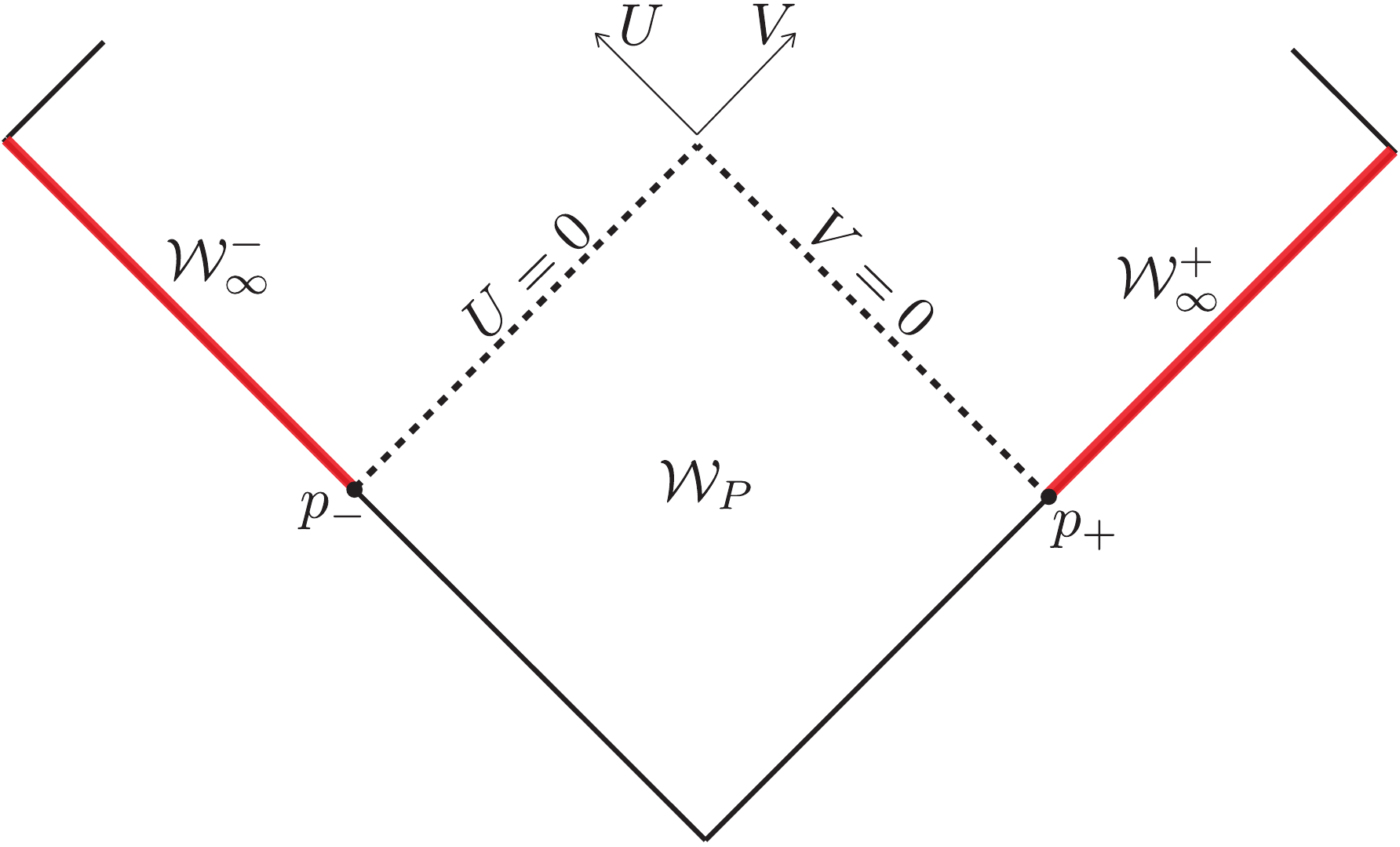} 
\end{minipage}
\caption{A demonstration of Eq.~\eqref{eq-assumption1}. The support of $\mathcal{D}(\mathcal{W}_\lambda^c)$ is on $\mathcal{W}^+_\lambda$ and $\mathcal{W}^-_\lambda$ which are sent to the past null infinity as $\lambda \to \infty$. The limiting region $\mathcal{W}^+_\infty$ ($\mathcal{W}^-_\infty$) is a future-inextendible generator of past null infinity with endpoint on $p_+$ ($p_-$). The leading singularity structure of a robust singularity $\langle \Phi^{(i)}(U,V,y^A)\rangle$ as $V\to 0^-$ is only sensitive to the presence of defect operators at $\mathcal{W}^+_\infty$. This significantly constrains the allowed robust singularity ansatz at $\partial \mathcal{W}_P$.}
\label{fig5}
\end{figure}

Now, note that $\mathcal{W}_\lambda^c = \mathcal{W}_\lambda^+ \cup \mathcal{W}_\lambda^-$, where for each $\lambda$, $\mathcal{W}_\lambda^\pm$ are the disjoint sets (See Fig. \ref{fig5}):
\begin{align}
&\mathcal{W}_\lambda^+ = \{V\geq0, U\leq -\lambda L \},\\
&\mathcal{W}_\lambda^- = \{V\leq-\lambda L, U\geq0 \}.
\end{align}
Let us write the following decomposition, in a notation where we set $\lambda = \infty$ to denote the limit:
\begin{align}\label{eq-n9865}
\mathcal{D}(\mathcal{W}_\infty^c) = \sum_{m_1, m_2 \in M} c_{m_1 m_2}~\Gamma^{(m_1)} (\mathcal{W}^+_\infty) \Gamma^{(m_2)} (\mathcal{W}^-_\infty)
\end{align}
where $c_{m_1 m_2}$ are complex numbers and $\Gamma^{(m_1)} (\mathcal{W}^+_\infty)$(respectively, $\Gamma^{(m_2)} (\mathcal{W}^-_\infty)$) are (in our new notation) indicating the $\lambda \to \infty$ limit of operator-valued distributions $\Gamma^{(m_1)} (\mathcal{W}^+_\lambda)$ ($\Gamma^{(m_2)} (\mathcal{W}^-_\lambda)$) with support contained in $\mathcal{W}^+_\lambda$($\mathcal{W}^-_\lambda$). The list $M$ generalizes the list $I$ of local operator-valued distributions (note the analogy between Eq.~\eqref{eq-12c5v} and Eq.~\eqref{eq-n9865} which can be proved generally). We have:
\begin{align}\label{eq-24c5v7b}
\langle \Phi^{(i)}(U,V,y^A) \rangle^{\mathcal{W}_P}_\omega = \sum_{m l} c_{m_1 m_2} \langle \Gamma^{(m_1)} (\mathcal{W}^+_\infty) \Gamma^{(m_2)} (\mathcal{W}^-_\infty)\Phi^{(i)}(U,V,y^A)  \rangle^{\mathbb{R}^{1,d-1}}_{\omega_0}
\end{align}
Since any state in Minkowski spacetime is by definition smooth at $\partial \mathcal{W}_P$, the origin of any 1-point function singularity on $V=0$ must be the presence of $\Gamma^{(m)}_+ (\infty)$ in Eq. \eqref{eq-24c5v7b}. Demanding that this singularity is robust ought to further constrain how the leading singularity arises from Eq.~\eqref{eq-24c5v7b}. Our argument for this involves extending (as an assumption) the following simple observation. Let $G(x)$ be defined through the following analogue of the correlator \eqref{eq-24c5v7b}:
\begin{align}
G(x) = \sum_{j_1,j_2\in I}c_{j_1,j_2}\langle \Phi^{(j_1)} (x_-) \Phi^{(j_2)} (x_+) \Phi^{(i)}(x)  \rangle^{\mathbb{R}^{1,d-1}}_{\omega_0},
\end{align}
Here, $G(x)$ is defined in $\mathbb{R}^{1,d-1} - J(x_+ \cup x_-)$, for $x_+ \neq x_-$ and $\omega_0$ is any state. We can analyze the limit $x \to x_+$ with the OPE (see Eq.~\ref{eq-OPE}):
\begin{align}
G(x)\stackrel{x\to x_+}{=}\sum_{j_1,j_2\in I}c_{j_1,j_2} \sum_{j_3 \in I}C^{(j_2)(i)}_{(j_3)}(x;x_+) \langle \Phi^{(j_1)} (x_-) \Phi^{(j_3)}(x_+) \rangle^{\mathbb{R}^{1,d-1}}_{\omega_0}
\end{align}
where $C^{(j_2)(i)}_{(j_3)}(x;x_+)$ are the OPE coefficients and (as in Eq.~\ref{eq-OPE}) the equality is to be understood as an asymptotic expansion in general. A priori, the leading singularity is given by a some subsets of indices $j_2 \in \tilde{I}_2 \subset I$ and $j_3 \in \tilde{I}_3 \subset I$ of the OPE coefficients $C^{(j_2)(i)}_{(j_3)}(x;x_+)$. Now, suppose we further demand that the leading singularity is the same for all states in the GNS Hilbert space built around $\omega_0$. The only way for this to happen is if $j_1 = j_3=\mathbb{1}$. It would then follow in particular that:
\begin{align}\label{eq-c3c2}
G(x) \stackrel{x\to x_+}{=} \sum_{j_2 \in \tilde{I}_2} c_{\mathbb{1},j_2} C^{(j_2)(i)}_{(\mathbb{1})}(x;x_+) + \cdots
\end{align}
where the ellipsis denotes subleading singularities. Our main takeaway from this analysis can now be summarized in the following equation:
\begin{align}\label{eq-35c35v}
\sum_{j_2 \in \tilde{I}_2} c_{\mathbb{1},j_2}\langle \Phi^{(j_2)} (x_+) \Phi^{(i)}(x) \rangle^{\mathbb{R}^{1,d-1}}_{\omega_0} \stackrel{x\to x_+}{=} \sum_{j_2 \in \tilde{I}_2} c_{\mathbb{1},j_2} C^{(j_2)(i)}_{(\mathbb{1})}(x;x_+) + \cdots
\end{align}
which can be trivially derived by applying the OPE to the LHS. This relation shows that a robust singularity's leading behavior in the $x \to x_+$ has a structure which is only sensitive to $\Phi^{(j_2)} (x_+)$.

One can extend the conclusion of Eq.~\eqref{eq-35c35v} to the case where $x$ approaches the light cone emanating from $x_+$ with an analogous argument using the so-called light cone OPE (see e.g.~\cite{Kravchuk:2018htv, Hartman:2016lgu}). And further to the case of a local operator approaching the light cone of the support of a defect operator (see e.g.~\cite{Billo:2016cpy}). Since neither of these well-studied objects are part of the axiomatic structure of~\cite{Hollands:2009bke}, here we will simply assume the analogous statement: the robustness of the $\langle \Phi^{(i)}(x)\rangle^{\mathcal{W}_P}_{\omega_0}$ singularity in Eq. \eqref{eq-35c35v} implies the following factorization
\begin{align}\label{eq-213cv7u3} 
\sum_{m_1 m_2} c_{m_1 m_2} \langle \Gamma^{(m_2)} (\mathcal{W}^-_\infty) \rangle^{\mathbb{R}^{1,d-1}}_{\omega_0} \langle \Gamma^{(m_1)} (\mathcal{W}^+_\infty) \Phi^{(i)}(U,V,y^A)\rangle^{\mathbb{R}^{1,d-1}}_{\omega_0} \stackrel{V\to0^-}{=} (P^{(i)}_{\omega})^{(0)}(U,V,y^A) + \cdots
\end{align}
A few comments about Eq.~\eqref{eq-213cv7u3} are in order. Firstly, as in Eq. \eqref{eq-c3c2} the leading singularity in Eq.~\eqref{eq-213cv7u3} may only be governed by a subset of the terms $m_2$, in which case the factorization of the summand according to Eq.~\eqref{eq-213cv7u3} is only implied for those $m_2$ values. However, since we only care about the leading singularity structure on the RHS this more strict condition is actually equivalent to \eqref{eq-213cv7u3}. Furthermore, Eq. \eqref{eq-35c35v} further requires that $\Gamma_-^{(l)}(\infty) = \mathbb{1}$ which is a special case of Eq.~\eqref{eq-213cv7u3}. In what follows, we use only this weaker condition.

Since all points in $\mathbb{R}^{1,d-1}$ with $V<0$ are spacelike separated from $\Gamma^{(m_1)} (\mathcal{W}^+_\infty)$, it follows that $\langle \Gamma^{(m_1)} (\mathcal{W}^+_\infty) \Phi^{(i)}(U,V,y^A)\rangle^{\mathbb{R}^{1,d-1}}_{\omega_0}$ are smooth in the $V<0$ region of Minkowski spacetime. Therefore, Eq.~\eqref{eq-213cv7u3} in particular implies that $(P^{(i)}_{\omega})^{(0)}(U,V,y^A)$ is smoothly extendible to $U=0$, i.e.,
\begin{align}\label{eq-23v2664v}
\lim_{U\to0^-}(P^{(i)}_{\omega})^{(0)}(U,V,y^A)=\text{finite}.
\end{align}

The condition~\eqref{eq-23v2664v} is a very strong constraint on the singularity structure of the $\mathcal{W}_P$ states. We will now exploit it in a Hilbert space containing a $k^\mu$-symmetric state $\omega$, i.e., symmetric under the flow of $k^\mu =  U\partial_U - V \partial_V$. The motivation for considering the symmetry is its presence in the Unruh and HH states of black holes. Let $\Phi_{\mu_1, \cdots, \mu_r}(x)$ be any operator with the indicated tensor structure which has a robust singularity at $V=0$.\footnote{Note that here we departure from the abstract notation $\Phi^{(i)}$ which does not make manifest the tensor structure of operators.}For example, $T_{VV}$ is a special case with $r=2$ and $\mu_1=\mu_2=V$. The stationarity of $\omega$ implies:
\begin{align}
\langle \Phi_{\mu_1, \cdots, \mu_r}(U,V,y^A) \rangle^{\mathcal{W}_P}_{\omega} =V^{-k}f^{(\mu_1, \cdots, \mu_r)}(\rho=UV,y^A).
\end{align}
where $k$ is the number of $V$ indices minus the number of $U$ indices in the specific component $\Phi_{\mu_1, \cdots, \mu_r}(x)$ under consideration, and the set of functions $f^{(\mu_1, \cdots, \mu_r)}(\rho,y^A)$, defined for $\rho>0$, are smooth and a priori allowed to diverge arbitrarily at $\rho = 0$. However, combined with Eq. \eqref{eq-23v2664v} they are constrained to satisfy
\begin{align}
f^{(\mu_1, \cdots, \mu_r)}(\rho,y^A) \stackrel{\rho \to 0^+}{=} o(\rho^{-\epsilon}),
\end{align}
where $\epsilon$ is any positive real number. Roughly speaking, any $f^{(\mu_1, \cdots, \mu_r)}(\rho,y^A)$ singularity at $\rho=0$ is constrained to be weaker than a power law.

The upshot is that for any $\Phi_{\mu_1, \cdots, \mu_r}(x)$ with a robust singularity, the leading \emph{allowed} singularity as $V \to 0^-$ satisfies:
\begin{align}
\langle \Phi_{\mu_1, \cdots, \mu_r}(U,V,y^A) \rangle^{\mathcal{W}_P}_{\omega} &\stackrel{V\to0^-}{=}O\left(\frac{1}{V^k}\right), \quad  \text{for  } k>0,\label{eq-mainmain1} \\ 
\langle \Phi_{\mu_1, \cdots, \mu_r}(U,V,y^A) \rangle^{\mathcal{W}_P}_{\omega} &\stackrel{V\to0^-}{=}o\left(\frac{1}{V^\epsilon}\right), \quad  \text{for  } k\leq0.\label{eq-mainmain2}
\end{align}
Importantly, our analysis does not guarantee that the leading allowed singularities will necessarily appear. However, in the absence of any additional symmetries or constraints one would expect them to manifest generically. As a consistency check, we show in Appendix~\ref{app444} that our example \eqref{eq-24v23vb77b} from Sec.~\ref{sec:33} for which $\langle \phi^2 \rangle^{\mathcal{W}_P}_\omega$ does not obey \eqref{eq-mainmain2} in fact is not a robust singularity.

The singularity structures in Eqs. \eqref{eq-mainmain1} and \eqref{eq-mainmain2} are reminiscent of the structures found numerically for a free scalar field $\phi$ in $d=4$ black holes for operators $\phi^2$ (corresponding to $r=0$)~\cite{Lanir:2018vgb}, $T_{VV}$ (corresponding to $r=2, k=2$)~\cite{Zilberman:2024jns, Hollands:2020qpe, Hollands:2019whz, Zilberman:2022aum}, and $T_{VA}$ (corresponding to $r=2, k=1$)~\cite{Klein:2024sdd} as discussed in the introduction. Moreover, it has been argued that the $T_{VV}$ and $T_{VA}$ singularities are robust by our definition~\cite{Hollands:2020qpe, Hintz:2023pak, Klein:2024sdd}.

Let us emphasize that in deriving the singularity structures \eqref{eq-mainmain1} and \eqref{eq-mainmain2} for any $k^\mu$-symmetric state, one needs only to assume Eq. \eqref{eq-213cv7u3} which is a much weaker condition than the robustness of the singularity. For example, if the singularity-generating defect operator has the structure,
\begin{align}
\mathcal{D}(\mathcal{W}^c_\lambda) = \frac{1}{2} \left(\mathcal{D}_1(\mathcal{W}^+_\lambda)+\mathcal{D}_2(\mathcal{W}^-_\lambda)\right),
\end{align}
it is easy to see that condition \eqref{eq-213cv7u3} follows, though the singularity may not be robust.

Given the relative weakness of condition Eq. \eqref{eq-213cv7u3}, it would be appealing to find the weakest physical condition on the $\mathcal{W}_P$ states $\omega$  which implies it. We offer two speculative directions to explore in this direction. One could consider weakening the robustness condition by allowing an arbitrary numerical coefficient on the RHS of Eq. \eqref{eq-h98g89y7} and attempt to show that Eq.~\eqref{eq-213cv7u3} follows as a consequence. Alternatively, instead of placing conditions related to robustness of the singularity, one can place a physically reasonable constraints on long distance correlations as two clusters of operators in an $n$-point function accumulate around $p_+$ and $p_-$ in $\mathcal{W}_P$ in a manner analogous to cluster decomposition (see Fig.~\ref{fig5}). We find it plausible that a suitable such constraint would in particular imply the factorized structure \eqref{eq-213cv7u3}. Explorations in these directions will be the subject of future work.

Finally, let us emphasize that the singularity ansatz in Eqs. \eqref{eq-mainmain1} and \eqref{eq-mainmain2} are contingent on our restriction to states in definition \eqref{def1}. As we explained in Sec.~\ref{sec:3}, regardless of considerations about the black hole interior, the QFT axioms naturally leads one to states of this type as example of wedge states with boundary singularities. Our evidence provided in subsection.~\ref{sec:41} for the generality of this construction, along with its success in reproducing outer horizon singularities, have led us to assume the same efficacy for deriving the relevant singularity structure on the Cauchy horizon (more precisely, the near Cauchy horizon limit $\mathcal{W}_P$). We find it reassuring that the mild singularity ansatz of black hole Cauchy horizon comes out as a natural consequence of our restriction to this class of states. That being said, to fully explain the mildness puzzle, we need to prove that the black hole Cauchy horizon states in the Unruh and HH states, in the strict $\mathcal{W}_P$ limit, are indeed defect constructible. One plausible scenario to this effect would be if indeed \emph{all} $\mathcal{W}_P$ states turn out to be defect constructible. Investigating this is beyond the scope of this work. Next we will describe the implication of the results of this section for black hole Cauchy horizon states by analyzing the near Cauchy horizon limit explicitly. We will also outline basic steps to directly apply the techniques here to a black hole Cauchy horizon, beyond just the strict $\mathcal{W}_P$ limit.

\section{Discussion}
\label{sec:5}

What do we learn about the singularity structure of a $d>2$ black hole spacetime $\mathcal{M}$ with Cauchy horizon $\mathcal{H}_C$, from the singularity structures~\eqref{eq-mainmain1} and \eqref{eq-mainmain2} we derived for $\mathcal{W}_P$? As we mentioned before, $\mathcal{W}_P$ is the universal near Cauchy horizon limit of \emph{all} black hole Cauchy horizons. Let us make this sharper. Note that any geometry $\mathcal{M}$ is characterized by a set of length scales. For example, take the $d=4$ RN geometry with metric
\begin{align}
ds^2 = -f(r) dt^2 + f(r)dr^2 + r^2 d\Omega_2^2,
\end{align}
in which 
\begin{align}\label{eq-3vb6}
f(r) = \left(1-\frac{r_+}{r}\right)\left(1-\frac{r_-}{r}\right),
\end{align}
where $r_+$ and $r_-$ are the outer and inner horizon radii respectively. In RN, the length scales characterizing the geometry are $r_+$ and $r_-$. Let $\ell$ be the smallest length scale of a general $\mathcal{M}$. All other length scales can be written in terms of $\ell$ multiplied by a list of dimensionless ratios $\{\beta_k\}$. For the RN example, $\ell=r_-$ and $\beta = r_+ / r_-$. Sufficiently close to $\mathcal{H}_C$, by holding $\{\beta_k\}$ fixed and taking $\ell$ to be large, the metric can be approximated as:
\begin{align}\label{eq-qwft}
ds^2 = -dUdV + \sum_{A=1}^{d-2} {(dy^A)}^2 + O\left( \left(\frac{UV}{\ell^2}\right)^p, \left(\frac{y^A}{\ell}\right)^q\right)
\end{align}
where $p$ and $q$ are positive real numbers. It is evident that by holding $(U,V,y^A)$ and the ratios $\{\beta_k\}$ fixed while taking $\ell \to \infty$, the metric \eqref{eq-qwft} becomes that of $\mathcal{W}_P$. From here on, by $\ell \to \infty$ we implicitly mean while holding $(U,V,y^A)$ and the ratios $\{\beta_k\}$ fixed. We can apply this limit to the set of all correlation functions and in doing so distill a $\mathcal{W}_P$ state from the near Cauchy horizon limit of a black hole Cauchy horizon. The resulting $\mathcal{W}_P$ state is then subject to our results in Sec.~\ref{sec:4}. Unfortunately, this limit could erase a lot of information about the singularity structure on $\mathcal{H}_C$. For example, suppose we have the following $\mathcal{H}_C$ singularity:
\begin{align}
\langle T_{VV} \rangle= \frac{a}{\ell^{\alpha}U^{\frac{d-2-\alpha}{2}}V^{\frac{d+2-\alpha}{2}}},
\end{align}
where $a$ is some dimensionless real constant and $\alpha>0$. The $\ell \to \infty$ limit would set this to zero, and therefore Eq. \eqref{eq-mainmain1} does not have any baring on it. Now, consider the following hypothetical singularity of $\langle T_{\mu\nu}\rangle$ on a general $\mathcal{H}_C$:
\begin{align}\label{eq-3v5235b}
\langle T_{\mu\nu} (U,V, y^A)\rangle =O\left(\frac{1}{U^{\frac{d-k}{2}} V^{\frac{d+k}{2}}}\right)
\end{align}
where $k$ is the number of $V$ indices minus the number of $U$ indices in the $T_{\mu\nu}$ component under consideration. This singularity does not on dimensional grounds allow any suppression by multiplicative factors of $\ell$, and therefore will survive the $\ell \to \infty$ limit. However, since none of the singularities in the ansatz \eqref{eq-3v5235b} are compatible with Eq. \eqref{eq-mainmain1} and \eqref{eq-mainmain2}, they are all ruled out as robust singularities. An obvious analogous statement can be made for other $1$-point function singularities that survive the $\ell \to \infty$ limit.

How do we extend the analysis of Sec.~\ref{sec:42} to a black hole Cauchy horizon directly, without taking the near Cauchy horizon limit? This would enable us to make statements about the mild singularity ansatz without taking the $\ell \to \infty$ limit which we have just shown erases information about the state. To do so, one would first need to come up with an analogue of definition~\ref{def1} for a black hole with a Cauchy horizon. This involves finding an analogue of $\mathcal{W}_0$ and its subset $\mathcal{W}$. Doing so might appear impossible at first since even though the black hole spacetimes are by definition extendible, they are not so necessarily into another globally hyperbolic spacetime. We do not expect this to be a serious obstacle though. The trick is to choose $\mathcal{W}$ to be only the black hole interior region (region II in right Fig.~\ref{fig1}). Let us describe how to find the appropriate $\mathcal{W}_0$ for such $\mathcal{W}$ in RN. We can consider an extension of this spacetime to $r<r_-$, in which we use an alternate smooth extension of $f(r)$ in $0\leq r<r_-$, so in particular $f(r=0)$ is finite (see e.g.~\cite{Hintz:2015jkj, Hollands:2019whz} where such a smooth extension is crucially used). This defines an extension of the black hole interior region $\mathcal{W}$ into a closed universe to the future of its $\mathcal{H}_C$. We can then choose $\mathcal{W}_0$ to be the spacetime region $0\leq r<r_+$ respectively, and pick $\mathcal{W}_\lambda$ exactly the same as in Eq.~\eqref{eq-b765} where $U$ and $V$ are now the inner Kruskal coordinates in RN. We expect that a similar construction would work in a Kerr-Newman black hole with a Cauchy horizon. This enables an analogue definition of defect constructible directly for a black hole interior. The analogue of the singularity structures Eq.~\eqref{eq-mainmain1} and \eqref{eq-mainmain2} would then immediately extend to $\mathcal{H}_C$.\footnote{This is of course subject to extending any assumptions in Sec.~\ref{sec:42} for $\mathcal{W}_P$ to this more general spacetime.} Indeed, the resulting mild singularity structure again remarkably matches the known numerical results. However, to make this a definitive derivation of the mild singularity structure from first principles of QFT, one would have to prove that the wide range of defect constructible states in $\mathcal{W}$ include the ones resulting from the HH or Unruh states. We leave this investigation to future work.

\section*{Acknowledgments}
I thank Raphael Bousso, Netta Engelhardt, Tom Faulkner, Luca Iliesiu, Adam Levine, Raghu Mahajan, Juan Maldacena, Marco Meineri, Geoff Penington, Pratik Rath, Steve Shenker, Jonathan Sorce, Douglas Stanford, and Robert Wald for discussions. I especially thank Raphael Bousso, Netta Engelhardt, Raghu Mahajan, Amos Ori, and Robert Wald for comments on an earlier draft of this paper. This work was supported by AFOSR through award FA9550-22-1-0098.

\appendix

\section{Review of the Microlocal Spectrum Condition}
\label{sec:app1}

Here we provide a brief review of the MSC discussed in Sec.~\ref{sec:3}. The review is directly based on~\cite{Hollands:2009bke} which discuss a particularly weak version of the constraint, and to which we refer the reader for a much more in depth discussion. The MSC can be phrased in terms of a constraint on the so-called wave front sets of QFT $n$-point functions.

The wave front in turn set can be defined generally for any distribution $u$ on $\mathbb{R}^d$. Let $\chi$ be a smooth function of compact support on $\mathbb{R}^d$. Then $\chi u$ is a distribution of compact support. Let $\hat{\chi u}$ denote its Fourier transform. The singular set of any compactly supported distribution such as $\hat{\chi u}$, denoted by $\Sigma(\hat{\chi u}(\lambda))$ is defined as the set of all $k \in \mathbb{R}^d$ such that
\begin{align}\label{eq-12c124}
|\hat{\chi u}(\lambda k)| \geq C \lambda^N,
\end{align}
for some $C>0$, and some $N$, and all $\lambda>0$. The wave front set of $u$ at a point $x$, denoted by $WF_x(u)$ is given by
\begin{align}
WF_x(u) = \cap_{\chi: x \in \text{support of } \chi} \Sigma (\chi u).
\end{align}
Looking at Eq.~\eqref{eq-12c124}, it is clear that the wave front set is empty at $x$ if and only if $u$ is smooth there. And a non-empty $WF_x(u)$, carries with it information about the ``direction'' of the non-smoothness in $u$. We can now define the wave front set of $u$ as
\begin{align}
WF(u) = \cup_{x\in \mathbb{R}^d} (x, WF_x(u)).
\end{align}
This definition is easily adaptable to arbitrary manifolds, like $\mathcal{W}$ where $WF(u)$ would be a subset of the cotangent bundle $T^* \mathcal{W}$. At each point $x$, a non-empty $WF_x(u)$ indicates directions of non-smoothness of $u$ at $x$.

Having defined the wave front set, the MSC as an axiom constraints the set of QFT $n$-point functions, as distributions on the manifold $\mathcal{W}^n$, to satisfy the property:
\begin{align}\label{eq-2v4v25}
WF(\langle \Phi^{(i_1)}(x_1) \cdots \Phi^{(i_n)}(x_n) \rangle^{\mathcal{W}}_{\omega}) \subset \Gamma_n(\mathcal{W}) \subset T^*\mathcal{W}^n
\end{align}
where $\Gamma_n(\mathcal{W})$ is a particular subset in the cotangent bundle, defined in the following multi-step way.

Let $g_{n,m}(\vec{x},\vec{y},\vec{p})$ denote an embedded graph with the following features: each graph has $n$ ``external vertices'' $x_1, \cdots, x_n \in \mathcal{W}$ and $m$ internal vertices $z_1, \cdots, z_m \in  \mathcal{W}$. The vertices can have arbitrary number of edges emanating from them. Each vertex $e$ corresponds to a null geodesics $\gamma_e : [0,1] \to \mathcal{W}$. There exists an ordering of vertices $x_1< \cdots< x_n$ and no such ordering for internal ones. If $e$ joins a source vertex $s(e)$ to a target vertex $t(e)$, then we have $s(e) = \gamma_e(0)$ and $t(e) = \gamma_e(1)$, and such that $e$ is directed from lower to higher order vertices. Each edge carries a future-directed tangent parallel covector field $\tilde{p}_e$. In particular, $\nabla_{\dot{\gamma}_e} \tilde{p}_e = 0$. We now define the following subset of $T^* \mathcal{W}^n$:

\begin{align}\label{eq-nij9d}
\Gamma_{n,m} (\mathcal{W}) = \{(x_1, p_1;\cdots:x_n,p_n)\in T^* \mathcal{W}^n - \{0\} | \exists~g_{n,m}(\vec{x},\vec{y},\vec{p})\nonumber\\
\text{such that } y_1,\cdots,y_k \in J^+(\{x_1, \cdots, x_n\}) \cap J^-(\{x_1, \cdots, x_n\})\nonumber\\
\text{such that } p_i = \sum_{e:s(e)=x_i} \tilde{p}_e - \sum_{e:t(e)=x_i} \tilde{p}_e \text{ for all $x_i$ and}\nonumber\\
\text{such that } 0 =\sum_{e:s(e)=y_i} \tilde{p}_e - \sum_{e:t(e)=y_i} \tilde{p}_e \text{ for all $y_i$} \}
\end{align}
where $J^{+}(.)$ ($J^{-}(.)$) denotes the causal future (past) of the set of points in its argument. The $\Gamma_n(\mathcal{W})$ in Eq.~\eqref{eq-2v4v25} is then defined as
\begin{align}
\Gamma_n(\mathcal{W}) =\overline{\cup_{m\geq0} \Gamma_{n,m}(\mathcal{W})}
\end{align}
where the over line denotes closure.

Now, suppose following Sec.~\ref{sec:32}, we define:
\begin{align}
\tilde{G}(y_1,\cdots, y_k, x_1, \cdots, x_n) = \langle \Phi^{j_1}(y_1)\cdots\Phi^{(j_k)}(y_k) \Phi^{i_1}(x_1)\cdots\Phi^{(i_n)}(x_n)   \rangle^{\mathcal{W}_0}_\omega
\end{align}
For any fixed $\Phi^{j_1}(y_1), \cdots, \Phi^{(j_k)}(y_k)$, let
\begin{align}
G(x_1, \cdots, x_n) = \langle \Phi^{j_1}(y_1)\cdots\Phi^{(j_k)}(y_k) \Phi^{i_1}(x_1)\cdots\Phi^{(i_n)}(x_n)   \rangle^{\mathcal{W}_0}_\omega
\end{align}
Note that $\tilde{G}$ is a distribution on $(\mathcal{W}_0)^{n+k}$ and $G$ is a distribution on $(\mathcal{W})^n$. The MSC implies:
\begin{align}
\tilde{G}(y_1,\cdots, y_k, x_1, \cdots, x_n) \subset \Gamma_{n+k}(\mathcal{W}_0)
\end{align}
It is now easy to see from Eq.~\eqref{eq-nij9d}, that if $y_1,\cdots, y_k \in \mathcal{W}^c$, the subset of $\mathcal{W}_0$ whose domain of influence does not intersect $\mathcal{W}$, then the following desirable result follows:
\begin{align}
G(x_1, \cdots, x_n) \subset \Gamma_n (\mathcal{W})
\end{align}
Therefore, $G(x_1, \cdots, x_n)$ is a suitable candidate for a new set of $n$-point functions for the spacetime region $\mathcal{W}$, at least as far as MSC is concerned.

\section{More Details on Defect Examples}
\label{sec:app3}
Here we review some technical details omitted from Sec.~\ref{sec:33}. First, let us review the definition of normal ordering, indicated by colons in the main text. For a general functional $\mathcal{F}(\phi(x))$:
\begin{align}\label{eq-24v2567n7}
:\mathcal{F}(\phi(x)):= \exp \left(dz_1 dz_2 \langle \phi(z_1) \phi(z_2)\rangle^{\mathbb{R}^{1,d-1}}_{\omega_0} \frac{\delta^2}{\delta \phi(z_1) \delta \phi(z_2)} \right) \mathcal{F}(\phi(x))
\end{align}
where the exponential on the RHS needs to be expanded before acting on $\mathcal{F}$ (see e.g.~\cite{Polchinski:1998rq}). For example, we have
\begin{align}
:\mathbb{1}: &= \mathbb{1},\\
:\phi(y): &= \phi(y).
\end{align}
If $\mathcal{F}$ contains products of $\phi(y)$ at the same point, a regularization procedure is implicit in Eq. \eqref{eq-24v2567n7}. For example, if $\mathcal{F} = \phi^2$, then we first separate the two $\phi$ operators, apply Eq. \eqref{eq-24v2567n7}, and then take the limit of the separation going to zero:
\begin{align}
:\phi^2(y): &= \left(1+ \int dz_1 dz_2 \langle \phi(z_1) \phi(z_2)\rangle^{\mathbb{R}^{1,d-1}}_{\omega_0} \frac{\delta^2}{\delta \phi(z_1) \delta \phi(z_2)} + \cdots )\right) \lim_{y' \to y} \phi(y) \phi(y')\nonumber\\
&=  \lim_{y' \to y} \left(\phi(y) \phi(y') - \langle \phi(y) \phi(y')\rangle^{\mathbb{R}^{1,d-1}}_{\omega_0}~\mathbb{1}\right).
\end{align}
This reproduces the familiar expression on the RHS for the normal ordered product $:\phi^2(y):$. We can also calculate normal ordering of a functional $\mathcal{F}$ involving $\phi$ operators at different locations. For example,
\begin{align}
:\phi(y_1) \phi(y_2): &= \phi(y_1) \phi(y_2) - \langle \phi(y_1) \phi(y_2)\rangle^{\mathbb{R}^{1,d-1}}_{\omega_0}~\mathbb{1}\\
:\phi^2(y_1) \phi^2(y_2): &= :\phi^2(y_1): :\phi^2(y_2): - 4 \langle \phi(y_1) \phi(y_2)\rangle^{\mathbb{R}^{1,d-1}}_{\omega_0}~\phi(y_1) \phi(y_2)\nonumber\\ 
&- \langle :\phi(y_1)^2: :\phi^2(y_2):\rangle^{\mathbb{R}^{1,d-1}}_{\omega_0}~\mathbb{1}
\end{align}
and so on. In effect, correlation functions involving $:\mathcal{F}(\phi):$ and other field operators can be computed by all wick contraction which do not involve contracting operators within $\mathcal{F}(\phi)$.

Now, we delve into some details of the defect examples in Sec.~\ref{sec:33}. Let us start by the defect \eqref{eq-124r31} in $\mathcal{W}_R$ for $d=4$. Deriving the $2$-point function is simple. Placed inside a correlator with two other operators $\phi(x_1)$ and $\phi(x_2)$, only the first two terms of \eqref{eq-124r31} yield non-zero answers:
\begin{align}
\langle \phi(x_1) \phi(x_2) \rangle^{\mathcal{W}_R}_\omega &= \langle \mathcal{D}(\mathcal{W}_R^c) \phi(x_1) \phi(x_2)\rangle^{\mathcal{W}_R}_{\omega_0}\nonumber\\
&= \langle \phi(x_1) \phi(x_2)\rangle^{\mathcal{W}_R}_{\omega_0} + \int_{\eth \mathcal{W}_R} d^2y~\langle :\phi^2(y): \phi(x_1) \phi(x_2)\rangle^{\mathcal{W}_R}_{\omega_0}
\end{align}

After the appropriate wick contractions we arrive at:
\begin{align}
\langle \phi(x_1) \phi(x_2) \rangle^{\mathcal{W}_R}_\omega = G_0(x_1,x_2) + 2 \int d^2 y~G_0(x_1,y) G_0(x_2, y)
\end{align}
which leads to Eq. \eqref{eq-4552} by explicit evaluation of the integral.

\section{The Non-robustness of the $\phi^2(x)$ Singularity in \eqref{eq-24v23vb77b} and its Robustness in \eqref{eq-2c1v55v6b}}
\label{app444}

The $\mathcal{W}_P$ state \eqref{eq-24v23vb77b} is Gaussian with $2$-point function:
\begin{align}
\langle \phi(x_1)\phi(x_2) \rangle^{\mathcal{W}_P}_\omega
=G_0(x_1, x_2) +\frac{1}{16 \pi^4 V_1 U_2} + \frac{1}{16 \pi^4 V_2 U_1}.
\end{align}
and results in the $1$-point function singularity:
\begin{align}
\langle \phi^2(x) \rangle^{\mathcal{W}_P}_\omega = \frac{1}{8\pi^4 UV} \implies (P^{\phi^2}_\omega)^{(0)}(U,V,y^A) = \frac{1}{8 \pi^4 U V}
\end{align}
which does not obey the structure in Eq. \eqref{eq-mainmain2} (for $r=0$). A prediction of our analysis in Sec.~\ref{sec:4} is that the singularity is not robust in the Hilbert space. Therefore, there ought to be another state $\omega'$ obtained from $\omega$ by acting with a member of the algebra of smeared fields in $\mathcal{W}_P$ such that
\begin{align}\label{eq-24vt22}
(P^{\phi^2}_{\omega'})^{(0)}(U,V,y^A) \neq \frac{1}{8 \pi^4 U V}.
\end{align}
We can construct this state by simply acting with a smeared field $\phi(f)$ where $f(x)$ is a real function with a compact support in $\mathcal{W}_P$. That is,
\begin{align}
\langle \phi^2(x) \rangle^{\mathcal{W}_P}_{\omega'} = \frac{\langle \phi(f) \phi^2(x) \phi(f) \rangle^{\mathcal{W}_P}_{\omega}}{\langle \phi(f)^2 \rangle^{\mathcal{W}_P}_{\omega}}
\end{align}
We can determine this by computing the 4-point function:
\begin{align}\label{eq-2c42c}
&\frac{\langle \phi(f) \phi(x_1) \phi(x_2) \phi(f) \rangle^{\mathcal{W}_P}_{\omega}}{\langle \phi(f)^2\rangle^{\mathcal{W}_P}_{\omega}} = \langle \phi(x_1) \phi(x_2) \rangle^{\mathcal{W}_P}_{\omega}\nonumber\\
&+ \frac{1}{\langle \phi(f)^2 \rangle^{\mathcal{W}_P}_{\omega}} \left(\langle \phi(f) \phi(x_1)\rangle^{\mathcal{W}_P}_{\omega} \langle \phi(f) \phi(x_2)\rangle^{\mathcal{W}_P}_{\omega} + \langle \phi(f)\phi(x_2)\rangle^{\mathcal{W}_P}_{\omega} \langle \phi(f) \phi(x_1)\rangle^{\mathcal{W}_P}_{\omega} \right)
\end{align}
We can then compute $\langle \phi^2(x) \rangle^{\mathcal{W}_P}_{\omega'}$ by subtracting $G_0(x_1,x_2)$ from the RHS of Eq. \eqref{eq-2c42c} and taking the coincident limit $x_2 \to x_1$:
\begin{align}\label{eq-12c35v}
\langle \phi^2(x) \rangle^{\mathcal{W}_P}_\omega &= \frac{1}{8\pi^4 U V} \nonumber\\
&+\frac{2}{\langle \phi(f)^2 \rangle^{\mathcal{W}_P}_{\omega}} \int d^4z_1 d^4z_2 f(z_1) f(z_2) \langle \phi(z_1) \phi(U,V,y^A)\rangle^{\mathcal{W}_P}_{\omega} \langle \phi(z_2) \phi(U,V,y^A)\rangle^{\mathcal{W}_P}_{\omega}
\end{align}
It is easy to see that the second term on the RHS of Eq. \eqref{eq-12c35v} scales as $1/V$ for a generic choice of $f(x)$, and therefore Eq.~\eqref{eq-24vt22} can be satisfied.

Next, consider the Gaussian state example in Eq.~\eqref{eq-4552} with the following $2$-point function:
\begin{align}\label{eq-24r23v}
\langle \phi(x_1)\phi(x_2) \rangle^{\mathcal{W}_R}_\omega
=G_0(x_1,x_2) +\frac{1}{16 \pi ^3}\frac{\log(\frac{U_1 V_1}{U_2 V_2})}{U_1 V_1 - U_2 V_2}.
\end{align}
By inspecting the defect operator in Eq.~\eqref{eq-124r31}, we can see that it is a conformal defect~\cite{Billo:2016cpy}. Furthermore, since the Minkowski vacuum expectation value of $\phi^2$ in the presence of the defect is non-zero, there must be an identity contribution to the defect - $\phi^2$ OPE, which by dimensional analysis (given the conformal nature of the defect) must be the sole leading singularity of the OPE. Therefore, $\phi^2$ is robust in the sector \eqref{eq-4552}. The conclusion is that any state in this sector must satisfy:
\begin{align}\label{eq-1235611x}
\langle \phi^2(x) \rangle^{\mathcal{W}_P}_\omega \stackrel{V\to0}{=} \frac{1}{16\pi^3 UV} +\cdots.
\end{align}
Therefore, a consistency check is to see that the change of state by acting with $\phi(f)$ (analogously to the previous example) does not change the leading singularity structure of Eq. \eqref{eq-1235611x}. This is evident by plugging the $2$-point function \eqref{eq-24r23v} to the RHS of Eq. \eqref{eq-12c35v}

\bibliographystyle{jhep}
\bibliography{all,all1}

\end{document}